\documentclass[pra,aps,twocolumn,showpacs,10pt]{revtex4-1}
\usepackage{amsmath,amssymb,graphicx,amsthm,dsfont,bm,color,ulem}

\begin{document}

\title{Time-diffracting beams: On their nature, diffraction-free propagation as needles of light, and nonlinear generation}

\author{Miguel A. Porras}

\affiliation{Grupo de Sistemas Complejos, ETSIME, Universidad Polit\'ecnica de Madrid, Rios Rosas 21, 28003 Madrid, Spain}

\begin{abstract}
We investigate on the properties of the recently introduced time-diffracting (TD) beams in free space. They are shown to be paraxial and quasi-monochromatic realizations of localized waves, spatiotemporal localized waves travelling undistorted at arbitrary speeds. The paraxial and quasi-monochromatic regime is shown to be necessary to observe what can properly be named diffraction in time. TD beams of finite energy travelling at quasi-luminal velocities are seen to form substantially longer foci or needles of light than the so-called abruptly focusing and defocusing needle of light, or limiting TD beam of infinite speed. Exploring the properties of TD beams under Lorentz transformations and transformation by paraxial optical systems, we realize that the relativistically moving nonlinear polarization of material media induced by a strongly localized fundamental pump wave generates a TD beam at its second harmonic, whose diffraction-free behavior as a needle of light in free space can be optimized with a standard $4f$-imager system.
\end{abstract}


\maketitle

\section{Introduction}

The idea of diffraction in time was originally introduced more than half a century ago \cite{MOSHINSKY} in the field of the quantum mechanics as a way to express that the localized wave function of a free particle spreads in time as a transversally localized light source does longitudinally in Fresnel diffraction, since the two phenomena a formally ruled by the same law, the Sch\"odinger equation. It is only very recently that diffraction in time expresses not only an analogy, but a distinct phenomenon in the propagation of a pulsed light beam, that is, of a wave packet localized in space and time \cite{KAMINER,PORRAS1,KONDAKCI3}. A time-diffracting beam is a pulsed light beam whose natural diffraction spreading is swapped from the longitudinal propagation distance to time. Diffractive focusing and subsequent defocusing, as in a Gaussian beam, occurs in time as the pulsed beam surpasses a fixed longitudinal location, and this behavior is the same at any longitudinal location, except for a delay in time. \cite{PORRAS1}

Time-diffracting (TD) beam are then diffraction-free waves that can have, in contrast to Bessel and Airy beams \cite{BESSEL,AIRY}, strongly localized transversal profiles. Their temporal localization is however weak, thus carrying finite instantaneous power but infinite energy. As diffraction-free pulsed beams, TD beams belong to the class of the so-called {\it localized waves} \cite{LOCALIZED,SAARI,ZIOLKOWSKI,RODRIGUEZ,BESIERIS,SALO,RECAMI2}. For diffraction-free propagation at (positive or negative) velocity $v_g$ along the $z$ direction, localized waves can be described as coherent superpositions of monochromatic plane wave (MPW) constituents of frequencies $\omega$ and wavevectors $\mathbf{k}=(\mathbf{k}_\perp, k_z)$ in the light cone, $(|\mathbf{k}_\perp|^2+ k_z^2)^{1/2}=\omega/c$, whose longitudinal components and frequencies are moreover linearly linked by $k_z=a+\omega/v_g$. The peculiarities that determines time-diffracting structure are: (i) the plane $k_z=a+\omega/v_g$ intersects the light cone at a positive (optical) frequency in the $k_z>0$ region (see Fig. \ref{Fig1}), and (ii) only MPW constituents about the vertex of the intersecting hyperbola, ellipse or parabola with sufficiently small transversal component of the wave vector, $\mathbf{k}_\perp\equiv (k_x,k_y)$, and therefore with frequencies close to the vertex frequency $\omega_0$ are excited, so that any of these conical sections can be approached by a parabola.

TD behavior was first suggested in relation to abruptly focusing and defocusing needles of light \cite{KAMINER}, where this phenomenon happens {\it simultaneously} at all axial positions in the needle of light. The spatiotemporal (ST) frequency correlations needed to attain this behavior are those of a vertical hyperbola in Fig. \ref{Fig1}(a) when $v_g=\infty$, were theoretically studied theoretically in \cite{KONDAKCI} and \cite{ALONSO}, and realized experimentally in \cite{KONDAKCI3} using spatial-beam modulation and ultrafast pulse-shaping techniques. In \cite{KONDAKCI3}, diffraction in time is dramatically demonstrated with an Airy wave packet that bends in time and not longitudinally.

General TD beams travelling at arbitrary velocities $v_g$ [arbitrarily tilted planes in Fig. \ref{Fig1}(a)] have been introduced in \cite{PORRAS1}, and exemplified with TD Gaussian beams. With paraxial and quasi-monochromatic (P\&QM) excitation about the vertex of the conic section, Fresnel diffraction of the transversal profile occurs in time, with a time delay from one to another transversal plane that depends on their distance and wave velocity. The idea of changing the wave velocity is also in \cite{KONDAKCI2}, where these diffraction-free pulsed beams are moreover realized experimentally, although without explicit mention to the time-diffracting property. The sheets of light produced in \cite{KONDAKCI2}, and the ST Airy wave packets in \cite{KONDAKCI3} involve spectral bandwidths $\Delta \lambda$ much smaller than the carrier wavelength $\lambda_0$, and transversal frequencies $\Delta k_x$ much smaller than the propagation constant $k_0$, and are therefore well-within the P\&QM regime of pulsed beam propagation.

In this paper we investigate in depth on the nature of TD beams, including their propagation and transformation by optical systems. We explore the limits to the length of the needles of light formed by these beams when they carry finite energy, and propose alternate methods of generation based on nonlinear optical phenomena.

We first explicitly derive the general expression of TD beams as the localized waves that verify conditions (i) and (ii) above. It comes out from our analysis that {\it the P\&QM condition in (ii) is not only a simplifying assumption, but is a necessary condition to observe what can properly qualified as diffraction in time}. TD beams with finite-energy are also strongly localized in time, and their finite diffraction-free distance, or propagation distance as a needle of light, is determined by the available spectral resolution or uncertainty in the ST frequency parabolic correlations (as fixed by the experimental set-ups \cite{KONDAKCI3,KONDAKCI2}) and by the TD velocity. With the same resolution, the length of the needles of light of TD beams with only slightly superluminal or subluminal velocity are seen to be substantially larger than the abruptly focusing and defocusing needle of light \cite{KAMINER} of the same width. The number of times that TD beams of certain width can beat the standard diffraction length is ultimately limited by the quotient between its carrier frequency and the spectral resolution, a number that can reach several tens of thousands in practice.

As localized waves, TD beams belong to three broad families: subluminal, luminal (focus wave modes), and superluminal. According to \cite{SAARI}, all members of the same family can be obtained through Lorentz transformations from three ``seed" TD beams belonging to each family. In particular subluminal TD beams are ``seeded" by moving monochromatic light beams, which points to a possibility to generate the subluminal family, at least, using moving sources, as suggested also in \cite{SAARI,KONDAKCI3}. The P\&QM regime opens new perspectives in the study of the propagation and transformation of localized waves by optical systems. As paraxial waves, this analysis can be carried out using standard methods as Fourier optics \cite{GOODMAN}. In relation to the generation of TD beams, we find that a standard $4f$ system, or double optical Fourier transform, images a TD beam to another TD beam whose velocity is controlled by the magnification of the system.

With all this in mind, we recall the basic fact in nonlinear optics that the nonlinear polarization in material media created by a pump wave acts as a subluminal or superluminal source (in the medium) of waves with other frequencies. Indeed, if the nonlinear polarization is induced by a strongly localized pump wave in space and time, ST frequency correlations in the form of hyperbolic, elliptic and parabolic ST spectra are well-known to arise spontaneously in a variety of nonlinear phenomena such as second harmonic generation \cite{CONTI,CONTI2,VALIULIS}, Kerr-induced instability in ultrashort pulses \cite{FACCIO}, cross-phase modulation \cite{AVERCHI}, and others \cite{LONGHI}. These couplings arise generally as a result of the more efficient amplification of MPW constituents that are phase matched to the driving pump pulse, and compete with the ST couplings needed for stationary propagation in the medium \cite{SAARI2,PORRAS3,PORRAS2}. In the filamentation of ultrashort pulses, for example, these ST frequency couplings manifest themselves as the phenomenon of conical emission \cite{FACCIO2}. It is also recognized that these ST frequency couplings promote the excitation of localized waves, generally referred to as X-waves, {\it in the nonlinear medium}. However, except in \cite{AVERCHI}, their diffraction-free property {\it in free space} beyond the nonlinear medium has not been tested.

Following the research line in \cite{CONTI}, we investigate numerically and analytically on the nature of the second harmonic localized wave generated under group-mismatched-dominated conditions by a strongly localized fundamental pump wave, and demonstrate that this localized wave is actually a TD beam. In free space, however, it does not exhibit any diffraction-free behavior, spreading immediately after the medium. A $4f$ system, nevertheless, images it to a diffraction-free TD beam in the form of a long needle of light beating its natural diffraction distance a number of times easily controllable by the length of the nonlinear crystal, the group mismatch and the pump duration. All signatures of actual diffraction in time, and in particular a {\it Gouy's phase shift in time}, are observed in this wave.

\section{Time-diffracting beams as paraxial and quasi-monochromatic localized waves}

\begin{figure}[!]
\centering
\includegraphics*[height=3.4cm]{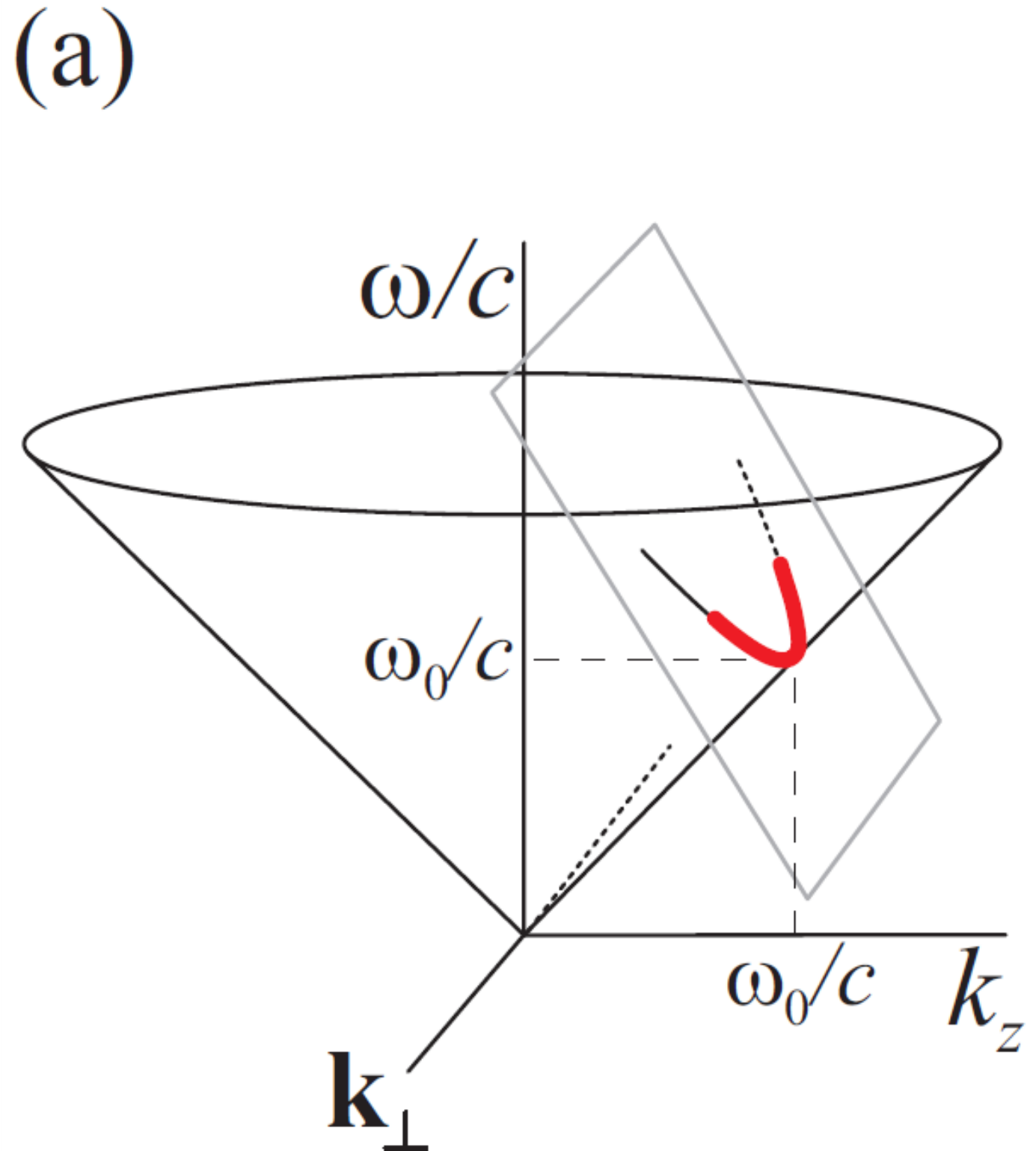}\includegraphics*[height=3.4cm]{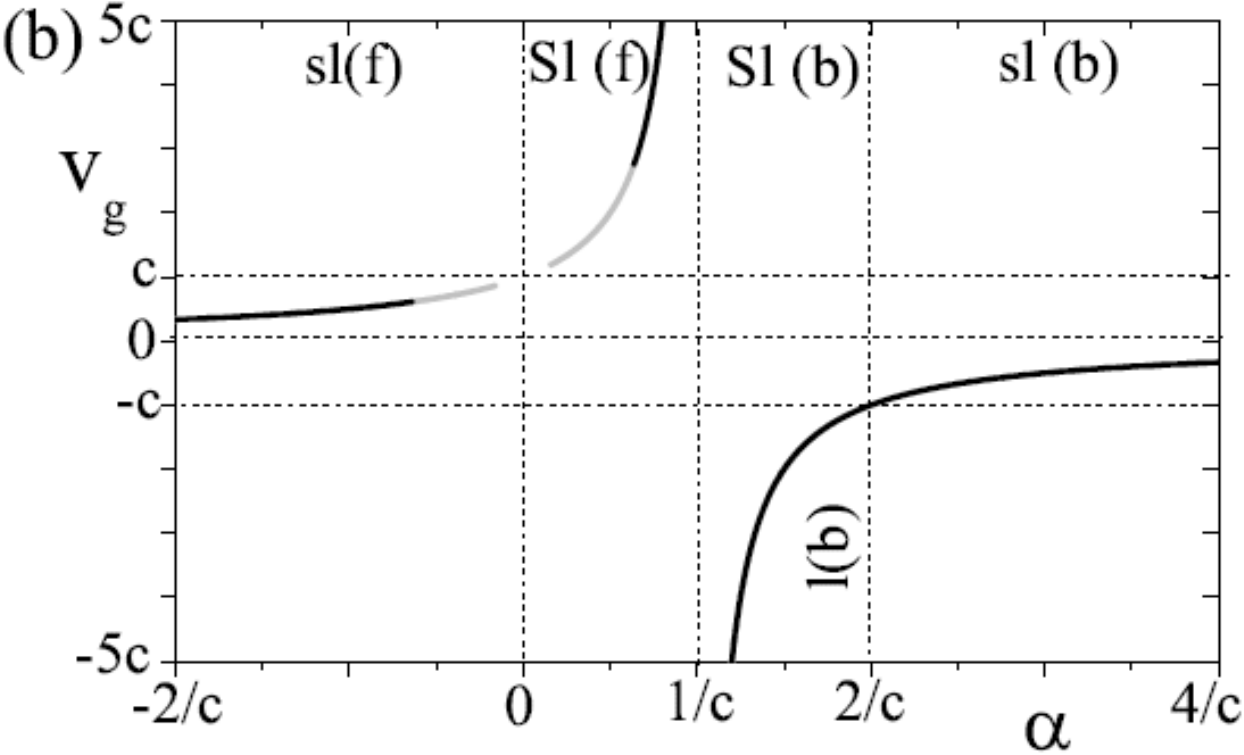}
\caption{\label{Fig1} (a) Intersection of the light-cone $(|\mathbf{k_\perp}|^2+k_z^2)^{1/2}=\omega/c$ with a plane $k_z=a+\omega/v_g$ at a positive frecuency $\omega_0$ in $k_z>0$. The excited MPW constituents (red curve) have small transversal wave vectors $\mathbf{k_\perp}$ and frequencies $\omega$ close to $\omega_0$. (b) Group delays $\alpha$ and corresponding group velocities $v_g$ of TD beams in the range of values of $\alpha$ satisfying the condition of quasi-monochromaticity (\ref{COND}). As a criterion, we set $\Delta\Omega< 0.1\omega_0$ (about $4$ cycle pulses in $\Delta t$) as the quasi-monochromatic limit. The black curve is for the quite extreme situation of a beam with only two wavelengths in its full width ($w_0=\lambda_0=2\pi c/\omega_0$), and the gray curve with four wavelengths in its full width ($w_0=2\lambda_0$). The gap about $\alpha=0$ is barely distinguishable for wider beams at the scale of the figure. The dashed vertical curves separate the ranges of values of $\alpha$ of subluminal (sl),  superluminal (Sl), forward (f) and backward (b) TD beams. The luminal (l) backward focus wave mode corresponds to $\alpha=2/c$.}
\end{figure}

As said, TD beams are paraxial, many-cycle realizations of localized waves with $k_z=a+\omega/v_g$ intersecting the light cone at a positive frequency $\omega_0$ in $k_z>0$. This condition implies that $k_{z,0}=a+\omega_0/v_g =k_0=\omega_0/c$, and the diffraction-free condition can be rewritten as $k_z=k_0+(\omega-\omega_0)/v_g$. The fundamental parameter defining the properties of a TD beam is its group delay with respect to a plane pulse  travelling at velocity $c$, defined as
\begin{equation}\label{ALPHA}
\alpha=\frac{1}{c}- \frac{1}{v_g}\neq 0  ,
\end{equation}
whereby the group velocity can be determined from $\alpha$ as $v_g=c/(1-\alpha c)$, as seen in Fig. \ref{Fig1}(b). The case $\alpha=0$ ($v_g=+c$) is excluded since it would yield a plane pulse, which is not a localized wave.

We write a generic pulsed beam solution to the wave equation for the electric field,
\begin{equation}\label{WE}
\Delta E -\frac{1}{c^2}\frac{\partial^2 E}{\partial t^2}=0,
\end{equation}
as a superposition of of forward propagating MPWs
\begin{equation}\label{E}
E(\mathbf{r}_\perp,t,z) = \int_{\omega>0} \!\!\!\!d\omega \!\int_{|\mathbf{k}_\perp|<\omega/c} \!\!\!\!\!\!\!d\mathbf{k_\perp} \hat E(\mathbf{k}_\perp,\omega) e^{ik_z z}e^{i\mathbf{k}_\perp\cdot \mathbf{r}_\perp} e^{-i\omega t},
\end{equation}
where $\mathbf{r}_\perp\equiv (x,y)$ and $k_z=\sqrt{(\omega/c)^2-|\mathbf{k}_\perp|^2}$. We assume that all MPW components travel at small angles to approach $k_z\simeq(\omega/c)-|\mathbf{k}_\perp|^2/2(\omega/c)$, and that all temporal frequencies are close to $\omega_0$ to approach $k_z\simeq \omega/c-|\mathbf{k}_\perp|^2/2k_0$. This amounts to neglect ST coupling effects {\it arising from propagation}, which are usually negligible for quasi-monochromatic, many-cycle pulses \cite{AKHMANOV,PORRAS4,PORRAS5}. With these approximations, we obtain an expression the electric field that can be expressed as the enveloped carrier oscillations $E(\mathbf{r}_\perp,t,z)=\psi(\mathbf{r}_\perp,t',z')e^{-i\omega_0 t'}$, with complex envelope
\begin{equation}\label{A1}
\psi(\mathbf{r}_\perp,t',z') = \int \!\!d\Omega\! \int \!\! d\mathbf{k_\perp} \hat E(\mathbf{k}_\perp,\Omega) e^{-i\frac{|\mathbf{k}_\perp|^2}{2k_0}z'}e^{i\mathbf{k}_\perp\cdot \mathbf{r}_\perp} e^{-i\Omega t'} ,
\end{equation}
and where we have introduced the local variables $t'=t-z/c$, $z'=z$, the detuning $\Omega=\omega-\omega_0$, and the integration limits are regarded as unnecessary for a strongly localized spectrum $\hat E(\mathbf{k}_\perp,\Omega)$ about $(\mathbf{k}_\perp,\Omega)=(0,0)$. For a factorized ST spectrum in the variables $\mathbf{k}_\perp$ and $\Omega$, Eq. (\ref{A1}) describes pulsed propagation as uncoupled paraxial diffraction at the frequency $\omega_0$ and undistorted pulse shape travelling at $c$, since ST coupling effects arising on propagation are neglected in this regime. Instead, we choose a coupled ST frequency spectrum of the form $\hat E(\mathbf{k}_\perp,\omega)=\hat \psi(\mathbf{k}_\perp)\delta[(k_z-k_0-\Omega/v_g)/\alpha]$ for non-diffracting propagation (the factor $1/\alpha$ is introduced to simplify later expressions). In the P\&QM approximation, the above ST spectrum reads $\hat E(\mathbf k_\perp,\Omega)=\hat \psi(\mathbf{k}_\perp)\delta(\Omega-|\mathbf{k}_\perp|^2/2k_0\alpha)$, whose support is the parabola
\begin{equation}\label{PARABOLA}
\Omega=\frac{|\mathbf{k}_\perp|^2}{2k_0\alpha}.
\end{equation}
The transversal frequency spectrum $\hat \psi(\mathbf{k}_\perp)$ is supposed to determine the localized transversal profile
\begin{equation}\label{IFT2}
\psi(\mathbf{r}_\perp)=\int d\mathbf{k_\perp} \hat \psi(\mathbf{k}_\perp) e^{i\mathbf{k}_\perp\cdot \mathbf{r}_\perp}.
\end{equation}
Although this is not the usual definition of inverse two-dimensional Fourier transform, we use it to avoid many $2\pi$ factors in the important expressions. With this choice
\begin{equation}\label{FT2}
\hat \psi(\mathbf{k}_\perp)=\frac{1}{(2\pi)^2}\int d\mathbf{r_\perp} \psi(\mathbf{r}_\perp) e^{-i\mathbf{k}_\perp\cdot \mathbf{r}_\perp} .
\end{equation}
Performing the integral in $\Omega$ in Eq. (\ref{A1}), we obtain
\begin{equation}\label{TD}
\psi_\alpha(\mathbf{r}_\perp,t',z') = \int  d\mathbf{k_\perp} \hat \psi(\mathbf{k}_\perp) e^{-i\frac{|\mathbf{k}_\perp|^2}{2k_0\alpha}(t'+\alpha z')}e^{i\mathbf{k}_\perp\cdot \mathbf{r}_\perp},
\end{equation}
which is the expression of a general TD beam. In \cite{PORRAS1} this expression was directly obtained as a solution of the form $\psi_\alpha(\mathbf{r}_\perp,t',z')=\psi(\mathbf{r}_\perp, t'+\alpha z')$ to the wave equation
\begin{equation}\label{PWE}
\Delta_\perp \psi + 2i k_0\frac{\partial \psi}{\partial z'} =0
\end{equation}
in the P\&QM approximations, but the present derivation demonstrates that TD beams are a subset of the family of localized waves. Also, Eq. (\ref{TD}) can be seen as the solution of Eq. (\ref{PWE}) with a transversal frequency spectrum $\hat \psi(\mathbf{k}_\perp,t')=\hat \psi(\mathbf{k}_\perp)e^{-i(|\mathbf{k}_\perp|^2/2k_0\alpha) t'}$ that depends slowly on time. In this view, the diffraction-free behavior results from the simultaneous quadratic phase modulation with increasing propagation distance and time, but remaining constant in the characteristics $t'+\alpha z'=\mbox{const.}$, or $z-v_gt= \mbox{const.}$

To simplify the discussion we assume that $\psi(\mathbf{r}_\perp)$ has no significant phase modulation. Also, since the novelty compared to Airy and Bessel-like beams is the strong (e. g., exponential) transversal localization, we will have in mind strongly localized transversal profiles $A(\mathbf{r}_\perp)$ occupying a region of approximate radius or half-width $w_0$, whose transversal frequency spectrum occupies a region of half-width $\Delta k_\perp\simeq 2/w_0$. According to Eq. (\ref{TD}), at $z'=0$ the TD beam features a intra-pulse temporal dynamics equal to that of paraxial or Fresnel diffraction of the monochromatic beam of profile $\psi(\mathbf{r}_\perp)$, with a ``temporal waist" of width $w_0$ at $t'=0$, and with a ``temporal propagation constant" $k_0\alpha$. By analogy with the standard diffraction length $z_R\simeq k_0w_0^2/2$ and the confocal parameter, or depth of focus, $L_R\simeq k_0w_0^2$, the diffraction time is $k_0|\alpha|w_0^2/2$ and the temporal depth of focus is $\Delta t\simeq k_0|\alpha|w_0^2$, which characterizes the full duration of the TD beam. At other locations $z'$, the only change is a shift in time, with the temporal waist located at $t'=-\alpha z'$ as a result of the superluminal or subluminal velocity.

The relation $\Delta t\simeq k_0|\alpha|w_0^2$ is the same as the relation $\Delta \Omega \simeq \Delta k_\perp^2/2k_0|\alpha|$, with the identification $\Delta \Omega\simeq 2/\Delta t$, imposed by the parabolic support of the spectrum in Eq. (\ref{PARABOLA}), and giving the temporal frequency bandwidth needed for a profile with transversal frequencies in $\Delta k_\perp$ to propagate without diffraction. The condition of quasi-monochromaticity, $\Delta \Omega=\Delta k_\perp^2/2k_0|\alpha|\ll \omega_0$, to which we are restricted, imposes the limitation to group delays $|\alpha|\gg \Delta k_\perp^2/2k_0^2c$, or
\begin{equation}\label{COND}
|\alpha|\gg  \frac{2}{L_R\omega_0}.
\end{equation}
This condition excludes interval about $\alpha=0$, or of velocities close to $v_g=c$, that is quite narrow for any paraxial beam ($\Delta k_\perp\ll k_0$), even for beams as narrow as a few wavelengths, as illustrated in Fig. \ref{Fig1}(b). Thus, given $\psi(\mathbf{r}_\perp)$ satisfying $\Delta k_\perp\ll k_0$, localized waves with forward and backward superluminal group velocity [$2/L_R\omega_0<\alpha<2/c$], forward and backward subluminal group velocity [$\alpha<-2/L_R\omega_0$ and $\alpha>2/c$), and backward luminal velocity ($\alpha=2/c$), or focus wave mode, can be realized as P\&QM TD beams with the profile $\psi(\mathbf{r}_\perp)$, all them characterized by a temporal-transversal pattern equal to that of the axial-transversal diffraction pattern for the profile $\psi(\mathbf{r}_\perp)$ in the Fresnel approximation. Among the different families of localized waves \cite{SAARI}, the only one that do not adopt the form of a TD beam under P\&QM conditions are X-type waves, whose plane $k_z=\omega/v_g$ intersects the light cone through the origin $\omega=0$ \cite{SAARI}.

We consider three TD beams belonging to the three families as particularly relevant: 1) The limiting superluminal or abrupt TD beams ($\alpha=1/c$, $v_g=\infty$) are P\&QM realizations of the first ``TD beams" (see discussion below) described in \cite{KAMINER}. Their conical section in Fig. \ref{Fig1}(a) is a vertical hyperbola. 2) Focus wave modes $(\alpha=2/c$, $v_g=-c$) are the only TD beams satisfying {\it exactly} the wave equation \cite{SEZGINER}. The plane $k_z=k_0-\Omega/c$ crosses the light cone at -45$^\circ$, and the conical section is therefore a parabola. 3) Monochromatic light beams can be regarded, according to Eq. (\ref{TD}), as the limiting subluminal TD beams ($\alpha=\pm\infty$), or static TD beams ($v_g=0$). The conical section is a horizontal circle giving, accordingly, spectral bandwidth $\Delta\Omega=\Delta k_\perp/2k_0|\alpha|=0$.

We point out that these results are valid regardless of whether the number of transverse dimension is one or two. Also, the seemingly different limitation to the values of $\alpha$ in Ref. \cite{PORRAS1} was obtained from the requirement of paraxiality for a given bandwidth $\Delta \Omega\ll\omega_0$, while condition (\ref{COND}) is the requirement of quasi-monochromaticity with given $\Delta k_\perp$, since the transversal profile $\psi(\mathbf{r}_\perp)$ is chosen in the present description.

We stress that the P\&QM conditions are not here mere simplifying assumptions, but necessary conditions to observe what can conceptually be qualified as diffraction in time. Starting again with Eq. (\ref{E}) with $\hat E(\mathbf{k}_\perp,\omega)=\hat \psi(\mathbf{k}_\perp)\delta[(k_z-k_0-(\omega-\omega_0)/v_g)/\alpha]$ but not performing any approximation, the result is a diffraction-free pulsed beam whose structure in the temporal and transversal dimensions cannot be identified with the diffraction pattern in the longitudinal and transversal dimensions a monochromatic light beam. Limiting the discussion to the case with $v_g=\infty$, \cite{KAMINER} we would obtain \cite{ALONSO}
\begin{equation}\label{INFTY}
E(\mathbf{r}_\perp,t,z)= \int d\mathbf{k}_\perp \psi(\mathbf{k}_\perp) e^{-i\sqrt{k_0^2c^2+|\mathbf{k}_\perp|^2c^2}\,t}e^{i\mathbf{k}_\perp\cdot\mathbf{r}_\perp} e^{ik_0z},
\end{equation}
which is indeed a non-diffracting pulsed beam whose intensity has no axial dynamics, and whose temporal dynamics does not correspond with the axial dynamics of paraxial or nonparaxial diffraction, but is simply the peculiar temporal dynamics of the abruptly focusing and defocusing pulsed beam (it would be a nonparaxial diffraction in time with the interchange $t\leftrightarrows -z/c$ if there were a minus sign in front of $|\mathbf{k}_\perp|^2$). With $\Delta k_\perp\ll k_0$, however, we can approximate $\sqrt{k_0^2c^2+|\mathbf{k}_\perp|^2c^2}\simeq k_0c(1+|\mathbf{k}_\perp|^2/2k_0^2)$, and the pulsed beam in Eq. (\ref{INFTY}) acquires the ST structure of Fresnel diffraction in time with temporal propagation constant $k_0/c$, and with a bandwidth $\Delta\Omega=\Delta k_\perp^2c/2k_0\ll \omega_0$ of a quasi-monochromatic pulse, i. e., the electric field of the abrupt TD beam in Eq. (\ref{TD}) with $\alpha=1/c$. Although the numerical difference between Eq. (\ref{INFTY}) and its paraxial approximation might be minimal, Eq. (\ref{INFTY}) cannot be said to describe a diffraction pattern swapped to time.

This also follows from the quite limited space-time analogy in the full wave equation if P\&QM conditions are not invoked. The ansatz $E=A(\mathbf{r_\perp},t) e^{ik_0 z}$ for abruptly focusing and defocusing needles of light \cite{KAMINER} in the wave equation yields the Klein-Gordon equation $\partial^2 A/\partial t^2- c^2\Delta_\perp A + c^2k_0^2A=0$, which is not the same as the Helmholtz equation $\partial^2 B/\partial z^2+ \Delta_\perp B +k_0^2 B=0$ for the monochromatic ansatz $E=B(\mathbf{r}_\perp,z) e^{-i\omega_0 t}$. Only if we further write $A=\psi(\mathbf{r_\perp},t)e^{-i\omega_0 t}$ and assume that $\psi$ varies slowly in $t$ and $\mathbf{r}_\perp$, the paraxial wave equation in time, $\Delta_\perp \psi +2i(\omega_0/c^2)\partial \psi/\partial t=0$, describes adequately the ST structure of the envelope.

In this respect it is intriguing the fact that in the only situation where P\&QM conditions are not required ---the focus wave mode---, diffraction in time is Fresnel diffraction, even if the focus wave mode is highly nonparaxial.

\section{Time-diffracting beams of finite energy and needles of light}\label{FINITE}

As noted in \cite{PORRAS1}, the instantaneous power of TD beams is independent of time in the same way as the power of a monochromatic light beam is independent of propagation distance, and therefore the energy carried by TD beams is unbounded. Expressions of finite-energy TD beams featuring quasi-non-diffracting behavior can be obtained starting again with Eq. (\ref{A1}) but replacing Dirac's delta function in the ST spectrum with a narrow function about $\Omega=0$ of bandwidth $\Delta \Omega_e\ll \Delta \Omega$, i. e., the ST spectrum $\hat E(\mathbf{k}_\perp,\omega)=\hat \psi(\mathbf{k}_\perp)\hat f[(k_z-a-\omega/v_g)/\alpha]=\hat \psi(\mathbf{k}_\perp)\hat f(\Omega-|\mathbf{k}_\perp|^2/2k_0\alpha)$ is a narrow band about the parabola $\Omega=|\mathbf{k}_\perp|^2/2k_0\alpha$. The bandwidth $\Delta \Omega_e$ can be identified with the uncertainty or resolution with which the ST frequency correlations in the parabolic spectrum can be introduced in practice \cite{KONDAKCI3,KONDAKCI2}. Proceeding as above we now obtain
\begin{eqnarray}\label{FETD}
\psi(\mathbf{r}_\perp,t',z') &=& \!f(t')\!\int \! d\mathbf{k_\perp} \hat \psi(\mathbf{k}_\perp) e^{-i\frac{|\mathbf{k}_\perp|^2}{2k_0\alpha}(t'+\alpha z')}e^{i\mathbf{k}_\perp\cdot \mathbf{r}_\perp} \nonumber \\
&=&\!f(t')\psi_\alpha(\mathbf{r}_\perp,t',z'),
\end{eqnarray}
which is an ideal TD beam of velocity $v_g$ enveloped by the long pulse
\begin{equation}
f(t')= \int d\Omega \hat f(\Omega)e^{-i\Omega t'},
\end{equation}
of velocity $c$ and of duration $\Delta t_e\simeq 2/\Delta\Omega_e\gg \Delta t$. The instantaneous power is now given by $P(t')=(2\pi)^2 |f(t')|^2\int d\mathbf{k}_\perp|\hat \psi(\mathbf{k}_\perp)|^2$, and the energy ${\cal E}=\int dt' P(t')$ is finite for any square-integrable $f(t')$.

\begin{figure}[!]
\centering
\includegraphics*[height=4.1cm]{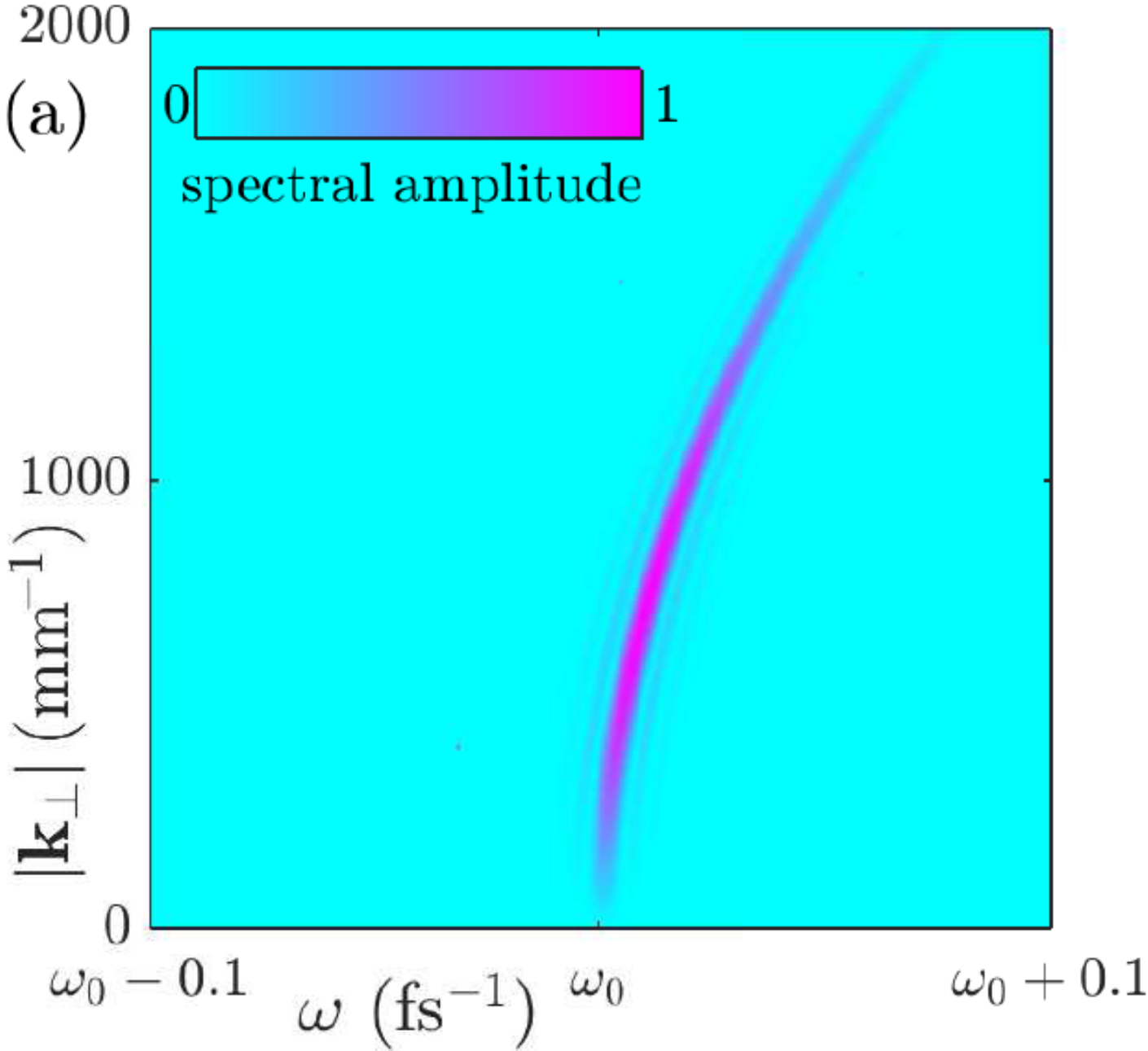}\includegraphics*[height=4.1cm]{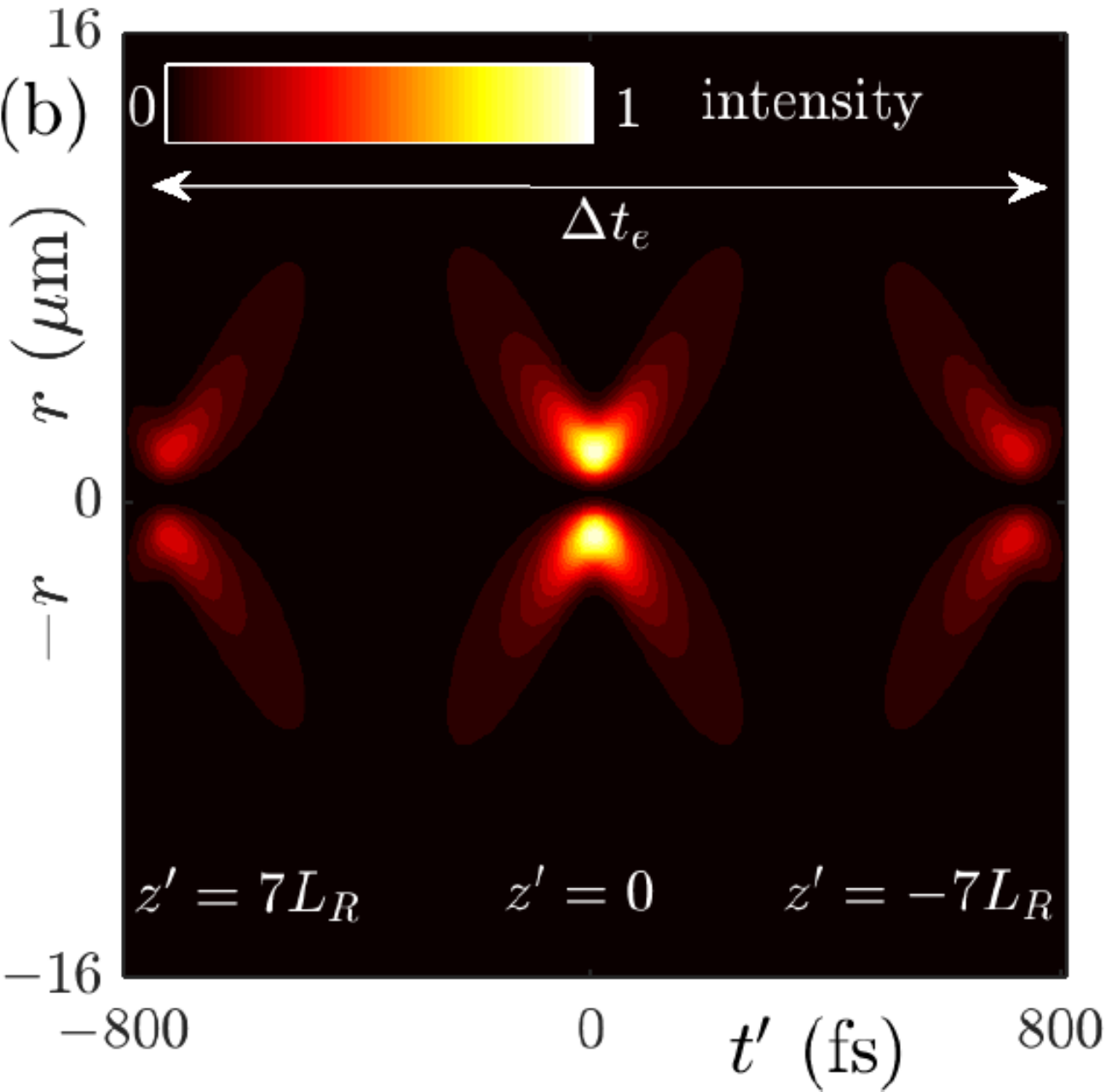}
\caption{\label{Fig2} Elegant Laguerre-Gauss TD beam with $n=0$, $s=1$, of carrier wavelength $\lambda_0=800$ nm, width $w_0=2\,\mu$m and group delay $\alpha=1/c$ ($v_g=\infty$). The envelope is $f(t')=e^{-t^{\prime 8}/\Delta t_e^8}$, of duration $\Delta t_e=1571$ fs, 15 times longer than $\Delta t=104.7$ fs. (a) ST spectral amplitude and (b) ST intensity profiles at several propagation distances. Temporal waist and envelope overlap from $z'=-7.5 L_R$ to $z'=7.5 L_R$, with $L_R=k_0w_0^2=0.031$ mm, i. e., $L_{\rm free}/L_R=15$.}
\end{figure}

As a few examples, we first take the Gaussian transversal frequency spectrum $\hat \psi(\mathbf{k}_\perp)=(4\pi/w_0^2) e^{-w_0^2|\mathbf{k}_\perp|^2/4}$ in Eq. (\ref{FETD}). The corresponding TD beam is the TD Gaussian beam
\begin{equation}\label{TDGB}
\psi(\mathbf{r}_\perp,t',z')=f(t')\frac{-iz_R}{q}\exp\left(\frac{ik_0|\mathbf{r}_\perp|^2}{2q}\right),
\end{equation}
where $z_R=k_0w_0^2/2$ and $q=(t'+\alpha z')/\alpha -i z_R$, and
expressed in \cite{PORRAS1} by an equivalent formula. Actually, it is not necessary to perform the integral in Eq. (\ref{FETD}) for every transversal frequency spectrum of interest: If the expression of a monochromatic light beam, $\psi(\mathbf{r}_\perp,z)$, is known in the literature, the replacement
\begin{equation}
z\rightarrow \frac{1}{\alpha}(t'+\alpha z')
\end{equation}
transforms it into a diffraction-free TD beam with the same transversal profiles in time, and multiplication by $f(t')$ into a finite-energy version of it. In this way it is straightforward to write down analytical expressions of diffraction-free (or quasi-diffraction-free) Hermite-Gauss, Leguerre-Gauss, Bessel-Gauss TD beams, or of the exponential-Airy TD beam synthesized in \cite{KONDAKCI3}. For example, the elegant Laguerre-Gauss beam \cite{PORRAS6}
\begin{equation}
\psi(\mathbf{r}_\perp,z)=\left(\frac{w_0^2}{w_C^2}\right)^{n+\frac{s}{2}+1}\left(\frac{r}{w_C}\right)^s L_n^s\left(\frac{r^2}{w_C^2}\right)e^{-\frac{r^2}{w_C^2}}e^{is\theta},
\end{equation}
with $w_C^2=w_0^2+2iz/k_0$, $n=0,1,\dots$, $s=0,\pm 1,\dots$, is transformed into the quasi-diffraction-free elegant Laguerre-Gauss TD beam with the redefinition $w_C^2=w_0^2+2i(t'+\alpha z')/k_0\alpha$, and multiplication by $f(t')$. For illustration purposes, Figs. \ref{Fig2}(a) and (b) show the ST spectral amplitude and the ST intensity profiles at several propagation distances of an elegant Laguerre-Gauss TD beam with $n=0$, $s=1$, carrier wavelength $\lambda_0=800$ nm, $w_0=2\,\mu$m and $\alpha=1/c\simeq 3333$ fs/mm ($v_g=\infty$). The TD duration is then $\Delta t = k_0|\alpha|w_0^2=104.7$ fs, corresponding to a bandwidth $\Delta\Omega = 0.019$ fs$^{-1}$. The envelope is taken to be a super-Gaussian pulse centered at $t'=0$ of duration $\Delta t_e=1571$ fs, so that the thickness or uncertainty in the parabolic spectrum is $\Delta\Omega_e = 0.0013$ fs$^{-1}$.

\begin{figure*}[!]
\centering
\includegraphics*[height=5.3cm]{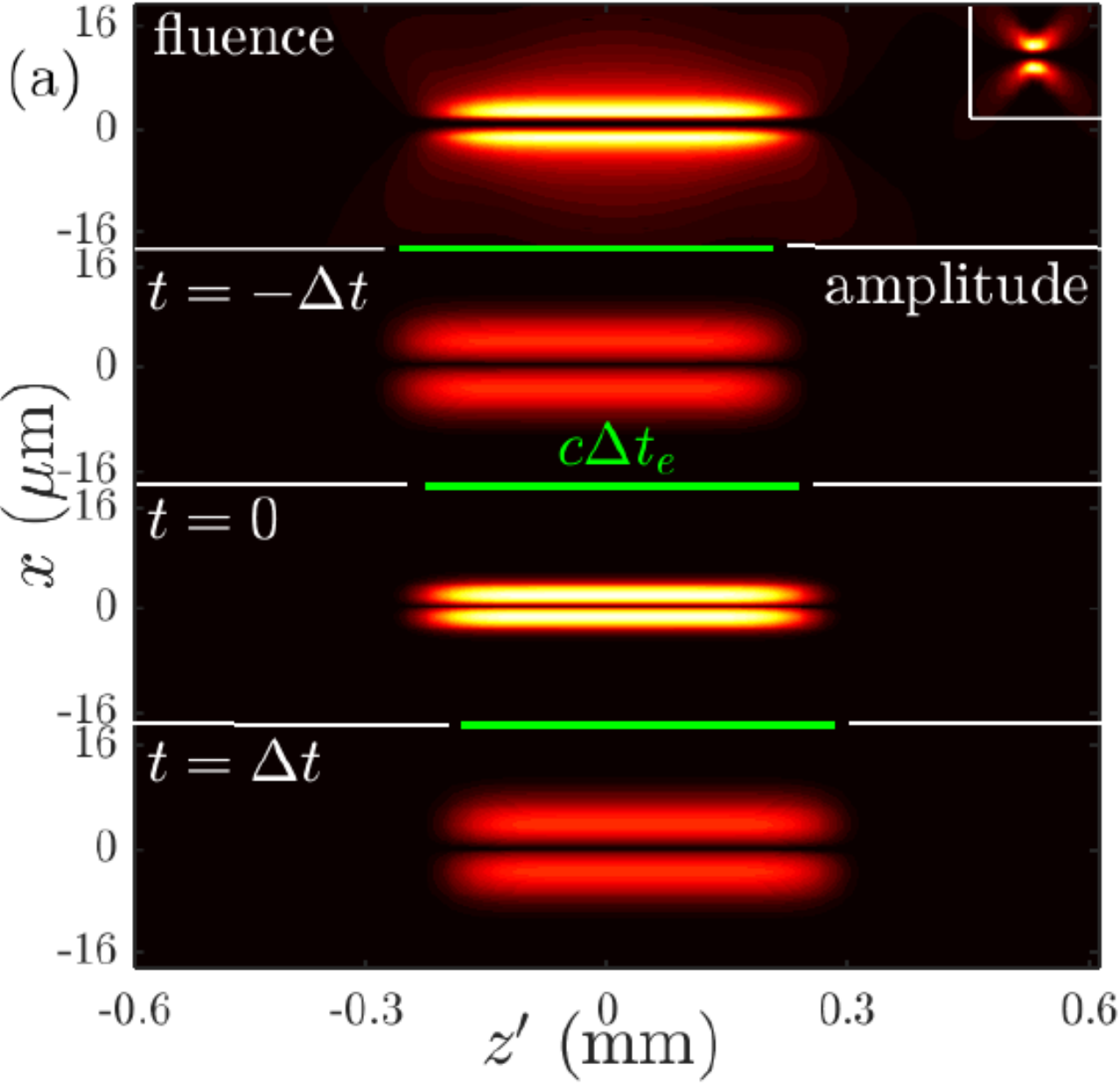}\hspace*{0.5cm}\includegraphics*[height=5.3cm]{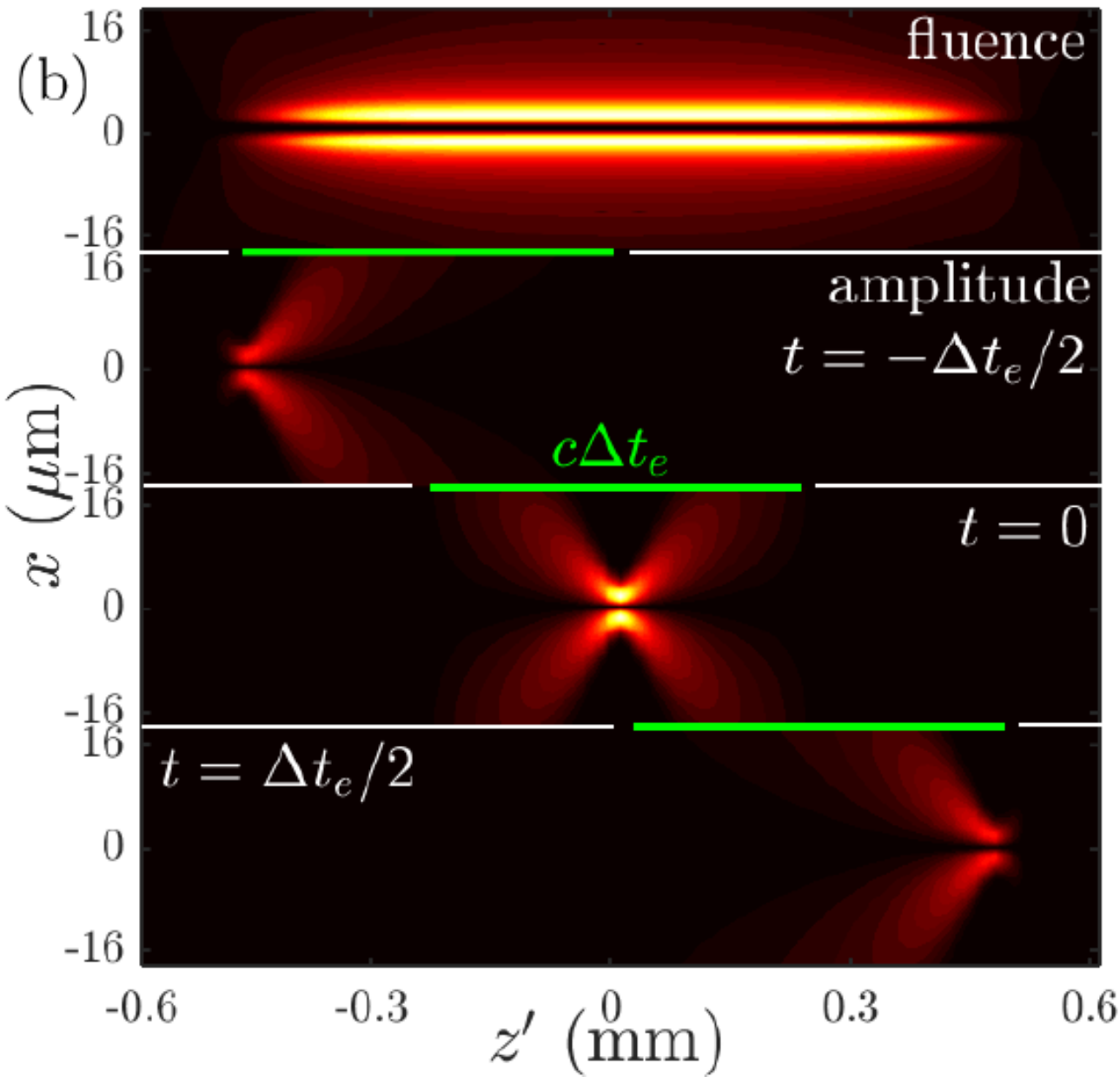}
\caption{\label{Fig3} Tubular fluence profiles of the TD beam in Fig. \ref{Fig2} (a) and of the TD beam of the same characteristics except that $\alpha=0.5/c$ (b). Inset: intensity profile of the original monochromatic elegant Laguerre-Gauss beam of the same width, represented using the same transversal and axial scales. The three lower panels in (a) and (b) represent snapshots of the amplitude at the indicated laboratory times $t$. The length of the thick green line is equal to the length $c\Delta t_e$ of the envelope, and translates from shot to shot at velocity $c$, while the TD beam translates at $v_g=\infty$ in (a), and at $v_g=2c$ in (b).}
\end{figure*}

The finite-energy TD beam will behave approximately as the ideal TD beam while its temporal waist and $f(t')$ overlap, as for the three distances in Fig. \ref{Fig2}(b). The walk-off distance $L_{\rm free}$ of the ideal TD beam and the luminal envelope is
given by $|\alpha|L_{\rm free}=\Delta t_e$, i. e., $L_{\rm free}=\Delta t_e/|\alpha|$. Compared to the standard confocal parameter $L_R=k_0w_0^2=\Delta t/|\alpha|$ of a monochromatic beam of the same width $w_0$, the finite-energy TD beam can be said to beat diffraction $L_{\rm free}/L_R=\Delta t_e/\Delta t$ times, or in terms of the spectral bandwidths, $\Delta \Omega/\Delta \Omega_e$ times. In the example of Fig. \ref{Fig2}, $L_{\rm free}=0.471$ mm, $L_{\rm free}/L_R=15$ times the confocal parameter. The beam fluence
\begin{equation}
F(\mathbf{r}_\perp,z') = \int dt' |\psi(\mathbf{r}_\perp,t',z')|^2 ,
\end{equation}
or energy per unit area of TD beams is shaped like a needle of light of length $L_{\rm free}$. For the elegant Laguerre-Gauss TD beam in Fig. \ref{Fig2}, the fluence actually forms the hollow needle of light of the length $L_{\rm free}/L_R=15$ shown in Fig. \ref{Fig3}(a, top panel). In the inset, the intensity profile of the monochromatic elegant Laguerre-Gauss beam of the same waist width is shown for comparison.

The needle shape was first mentioned in relation to the abrupt TD beam, but TD beams sinthetized with the same resolution and of the same width may be much longer. Given $\Delta \Omega_e$ and $\Delta k_\perp$, $L_{\rm free}/L_R=\Delta k_\perp^2/2k_0 |\alpha|\Delta\Omega_e$ is larger as $\alpha$ diminishes down to the limit (\ref{COND}) of quasi-monochromaticity, i. e., for slightly superluminal or subluminal TD beams, which imposes the upper bound $L_{\rm free}/L_R \ll \omega_0/\Delta\Omega_e$. In the example of Fig. \ref{Fig2}, the length $L_{\rm free}/L_R=15$ can be considerably improved below the limit $L_{\rm free}/L_R \ll 1850$ by simply diminishing $\alpha$. For instance, the only change from Fig. \ref{Fig3}(a) to (b) is that $\alpha=0.5/c\simeq 1667$ fs/mm ($v_g=2c$) instead of $1/c\simeq 3333$ (so that the TD duration $\Delta t=k_0|\alpha|w_0^2=52.35$ fs is reduced by half and its bandwidth $\Delta\Omega=0.038$ fs$^{-1}$ is double). With the same width spectral resolution, the length of the needle of light is $L_{\rm free}/L_R=30$. The reason of the longer needle of light can be visualized in the snapshots of the field amplitude in the laboratory time $t$ shown in the lower panels of Figs. \ref{Fig3}(a) and (b). The green horizontal segments indicate the location and length $c\Delta t_e$ of the luminal envelope at the three increasing instants of time. The abrupt TD beam appears and disappears simultaneously at all axial locations (its axial length is $v_g\Delta t=\infty$) within the envelope and in the short lapse of time $\Delta t$ during which the envelope does not move appreciably, giving a needle length $L_{\rm free}=\Delta t_e/|\alpha|= c\Delta t_e$ equal to the envelope length. With lower velocity ($\alpha<1/c$), the TD beam and its envelope overlap during a time during which the envelope advances appreciably, giving the longer needle length $L_{\rm free}=\Delta t_e/|\alpha|>c\Delta t_e$, twice longer in the example of the figure.

On the opposite side, the condition $L_{\rm free}/L_R= \Delta t_e/k_0w_0^2|\alpha|>1$ for appreciable quasi-diffraction-free behavior imposes the upper bound
\begin{equation}\label{ALPHAL}
|\alpha|< \frac{2}{L_R\Delta\Omega_e}
\end{equation}
to the group delays of TD beams of a given width and spectral resolution to form a needle of light, or $|\alpha|<15/c\simeq 50000$ in the above example. Indeed, for $\alpha$ surpassing that limit, $\Delta t=k_0|\alpha|w_0^2>\Delta t_e$, meaning that the temporal variation of the ideal TD beam $\psi_\alpha(\mathbf{r}_\perp,t',z')$ within the envelope $f(t')$ is increasingly negligible, and that the TD beam starts to behave as a standard (diffracting) pulsed beam $f(t')\psi(\mathbf{r}_\perp,z')$, whose fluence profile is proportional to the intensity profile of the original monochromatic light beam [as in the inset of Fig. \ref{Fig3}(a)].

In the above reasonings an unbounded space from $z'=-\infty$ to $+\infty$ is implicitly assumed (the ideal TD beam always crosses the whole envelope). In a limited space, e. g., a TD beam generated at $z'=0$ and existing only in $z'\ge0$ (as in Sec. \ref{NONLINEAR}), $L_{\rm free}$ is an upper bound to the effective diffraction-free distance, determined by the relative position of $f(t')$ and $\psi_\alpha$ upon generation. If for instance, the above elegant Laguerre-Gauss TD beams, with $f(t')$ and $\psi_\alpha$ initially ($z'=0$) centered at $t'=0$, exist only in $z'\ge 0$, the effective diffraction-free distance is $L_{\rm free}/2$.

\section{Transformation in moving reference frames}\label{FRAMES}

TD beams in with one transversal dimension have been synthesized in \cite{KONDAKCI3,KONDAKCI2} using sophisticated beam and pulse shaping techniques whose application to two transversal dimensions is not evident. Searching for alternative methods, we first recall the result in \cite{SAARI} that the different members each of the superluminal, subluminal, and luminal families of localized waves, are substantially one and the same localized wave observed in reference frames moving at different constant velocities. As a subset of localized waves, this property remains true for TD beams: Consider a reference frame using coordinates and time $(\mathbf{r}_{\perp,2},z_2,t_2)$ and moving at velocity $v$ ($|v|<c$) along the $z_1$ direction with respect to the ``laboratory" frame using parallel coordinate axes and time $(\mathbf{r}_{\perp,1},z_1,t_1)$. Using Lorentz transformations,
\begin{eqnarray}
\mathbf{r}_{\perp,2}&=&\mathbf{r}_{\perp,1},\\
z_2&=&\gamma(z_1-vt_1),\\
t_2&=&\gamma\left(t_1-\frac{v}{c^2}z_1\right),
\end{eqnarray}
with $\gamma=(1-v^2/c^2)^{-1/2}$, it is a straightforward calculation to verify that a P\&QM wave $E_2= \psi_2(\mathbf{r}_{\perp,2},t_2,z_2)e^{-i\omega_{0,2}t'_2}$ of carrier frequency $\omega_{0,2}$ and satisfying Eq. (\ref{PWE}) with $k_{0,2}$, transforms into $E_1=\psi_1(\mathbf{r}_{\perp,1},t_1,z_1)e^{-i\omega_{0,1}t'_1}$ of carrier frequency $\omega_{0,1}=\omega_{0,2}\gamma (1+v/c)$ and satisfying also Eq. (\ref{PWE}) with $k_{0,1}=k_{0,2}\gamma (1+v/c)$. In particular, a finite-energy TD beam of the form $f_2(t'_2)\psi_2(\mathbf{r}_{\perp,2},t'_2+\alpha_2 z'_2)$ transforms in to another TD beam $f_1(t'_1)\psi_1(\mathbf{r}_{\perp,1},t'_1+\alpha_1 z'_1)$ with
\begin{equation}\label{ALPHAT}
\alpha_1= \frac{\alpha_2(1-v/c)}{(1+v/c)-\alpha_2 v} .
\end{equation}
Of course, these mathematical relations must be taken with caution, since the wave in the laboratory frame satisfies formally Eq. (\ref{PWE}) but may not satisfy the P\&QM conditions, which sets some restrictions to the relative velocity, depending on the particular P\&QM wave in the moving frame. With these restrictions, inspection of Eq. (\ref{ALPHAT}) shows that each particular subluminal, superluminal or luminal TD beam transforms into other TD beams within the same family. In particular, an abrupt TD beam ($\alpha_2=1/c$, $v_{g,2}=\infty$) transforms into a superluminal TD beam with $\alpha_1=(1-v/c)/c$, or $v_{g,1}=c^2/v>c$a, a focus wave mode ($\alpha_2=2/c$, $v_{g,2}=-c$) is transformed into another focus wave ($\alpha_1=2/c$, $v_{g,1}=-c$), and a monochromatic light beam ($\alpha_2=\pm \infty$, $v_{g,2}=0$) does into a subluminal TD beam with $\alpha_{1}=1/c-1/v$, i. e., with group velocity $v_{g,1}=v$ equal to the velocity of the moving frame. The fact that a {\it diffracting monochromatic light beam} in a moving frame along the beam propagation direction is observed as a {\it diffraction-free pulsed light beam} in the laboratory frame seems obvious and striking at the same time. The same idea has recently been expressed in  \cite{KONDAKCI3} in the equivalent perspective that a monochromatic light beam in the laboratory frame would be observed as a diffraction-free pulsed beam by moving detectors. In Sec. \ref{NONLINEAR} we consider moving sources as a more practical approach to the generation of TD beams.

\section{Transformation by paraxial optical systems}\label{TRANSFORMATIONS}

P\&QM conditions allow for a quite simple analysis of the transformation of TD beam by optical systems. Among them, a simple $4f$-imager system, or two consecutive optical Fourier transforms, images a TD beam to another TD beam of different velocity.

Equation (\ref{FETD}) for ideal [$f(t')=1$] or for finite-energy TD beams can be expressed as
\begin{equation}\label{FEA}
\psi(\mathbf{r}_\perp, t',z')=  \int  d\mathbf{k_\perp} \hat \psi(\mathbf{k}_\perp,t') e^{-i\frac{|\mathbf{k}_\perp|^2}{2k_0} z'}e^{i\mathbf{k}_\perp\cdot \mathbf{r}_\perp},
\end{equation}
with
\begin{equation}\label{FEAT}
\hat \psi(\mathbf{k}_\perp,t')=f(t')\hat \psi_\alpha(\mathbf{k}_\perp,t')= f(t')\hat \psi(\mathbf{k}_\perp)e^{-i\frac{|\mathbf{k}_\perp|^2}{2k_0\alpha} t'} .
\end{equation}
The exponential $e^{-i(|\mathbf{k}_\perp|^2/2k_0) z'}$ in Eq. (\ref{FEA}) is the well-known propagator of the transversal frequency spectrum in the paraxial approximation, with a transversal frequency spectrum in Eq. (\ref{FEAT}) that depends slowly on time if $\Delta \Omega_e\ll\Delta\Omega\ll \omega_0$. Under these conditions, the transformation of TD beams by common optical systems is adequately described by the well-known rules of Fourier optics in the Fresnel approximation with a transversal frequency spectrum that depends slowly on the local time $t'$ at each particular axial location.

\begin{figure}[t]
\centering
\includegraphics*[height=2cm]{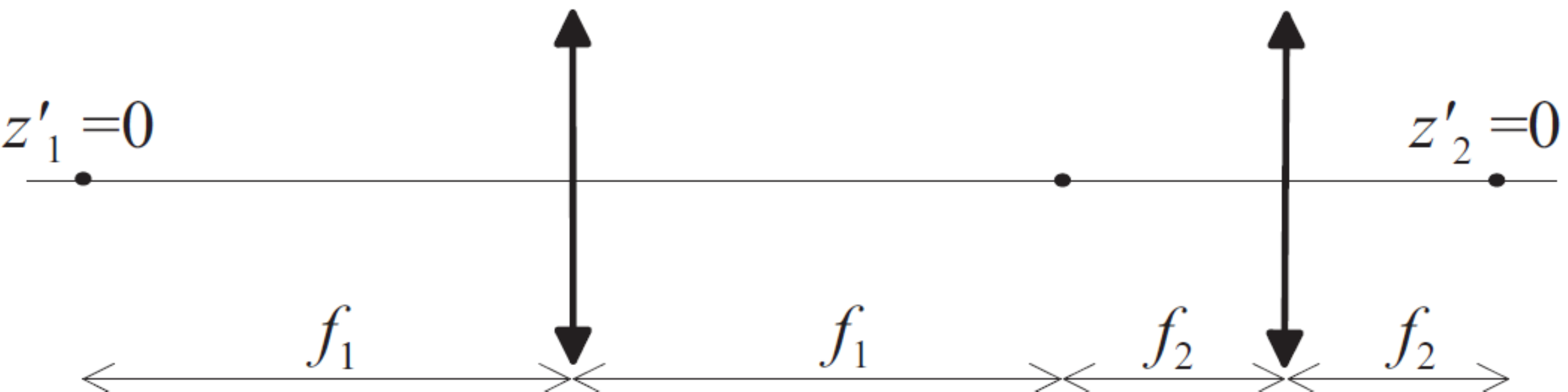}
\caption{\label{Fig4} $4f$ system and axial coordinates used in the analysis.}
\end{figure}

We find it particularly relevant the $4f$ system sketched in Fig. \ref{Fig4}, consisting preferably of two large enough spherical or parabolic mirrors (to avoid aperture effects, and chromatic and spherical aberrations) of focal lengths $f_1$ and $f_2$, and imaging $\psi_1(\mathbf{r}_\perp,t'_1)$ at the front focal plane into $\psi_2=(f_1/f_2)\psi_1[-(f_1/f_2)\mathbf{r}_\perp, t'_2]$, with $t'_2=t'_1-(2f_1+2f_2)/c$, at the back focal plane of the system. For the TD beam $\psi_1=f(t'_1)\psi_{\alpha_1}(\mathbf{r}_\perp,t'_1,z_1)$ in Eq. (\ref{FETD}) illuminating the system we can set, without loss of generality, $z'_1=0$ at the front focal plane of the $f_1$ mirror (Fig. \ref{Fig4}). Thus, the complex envelope at the back focal plane of the $f_2$ mirror will be
\begin{equation}
\psi_2(\mathbf{r}_\perp,t'_2)=\frac{f_1}{f_2}f(t'_2)\int d\mathbf{k}_\perp \hat \psi_1(\mathbf{k}_\perp)e^{\frac{-i|\mathbf{k}_\perp|^2}{2k_0\alpha}t'_2} e^{-i\mathbf{k}_\perp\cdot \frac{f_1}{f_2}\mathbf{r}_\perp} ,
\end{equation}
or, with the variable change $\mathbf{q}_\perp=-(f_1/f_2)\mathbf{k}_\perp$,
\begin{equation}
\psi_2(\mathbf{r}_\perp,t'_2)= \frac{f_1}{f_2}f(t'_2)\!\int \!\!d\mathbf{q}_\perp \hat \psi_1\!\left(\!-\frac{f_2}{f_1}\mathbf{q}_\perp\!\right)e^{\frac{-i|\mathbf{q}_\perp|^2}{2k_0\alpha_2}t'_2}e^{i\mathbf{q}_\perp\cdot \mathbf{r}_\perp},
\end{equation}
with
\begin{equation}\label{ALPHAN}
\alpha_2=\frac{f_1^2}{f_2^2}\alpha_1.
\end{equation}
Propagation a (positive or negative) distance $z'_2$ (Fig. \ref{Fig4}) about the back focus yields
\begin{eqnarray}
\psi_2(\mathbf{r}_\perp,t'_2,z'_2) &=& \frac{f_1}{f_2}f(t'_2)\int d\mathbf{q}_\perp \hat \psi_1\left(\!-\frac{f_2}{f_1}\mathbf{q}_\perp\!\right)
\nonumber \\
 &\times& e^{\frac{-i|\mathbf{q}_\perp|^2}{2k_0\alpha_2}(t'_2+\alpha_2 z'_2)}e^{i\mathbf{q}_\perp\cdot \mathbf{r}_\perp},
\end{eqnarray}
where $t'_2=t'_1-(2f_1+2f_2)/c-z'_2/c$. The $4f$ system then transforms a TD beam of group delay $\alpha_1$, width $w_{0,1}$ and envelope envelope $f(t'_1)$ into a compressed or expanded TD beam of of width $w_{0,2}= (f_2/f_1)w_{0,1}$, group delay $\alpha_2$ given by Eq. (\ref{ALPHAN}), and the same envelope $f(t'_2)$. Consequently the durations of the underlying ideal TD beam and of its envelope are the same as those of the input TD beam, the diffraction-free distance is scaled as $L_{{\rm free},2}=(f_2^2/f_1^2)L_{{\rm free},1}$, but the number of times that the output TD beam beats diffraction, $L_{{\rm free},2}/L_{R,2}$, is the same as that for the input TD beam. This result suggests that it would suffice to produce a single TD beam with positive $\alpha$ to produce all others, and the same for negative $\alpha$.

\section{Nonlinear generation of time-diffracting beams}\label{NONLINEAR}

Also, the analysis in Sec. \ref{FRAMES} suggests that a TD beam could be generated by a moving source of light, but this possibility appears to be limited, in free space, to subluminal velocities and therefore to subluminal TD beams. Nonlinear optics in material media offer, however, many examples of light acting as sources of light that move at relativistic velocities,  even at superluminal velocities {\it in the medium}. Among them, a well-studied example is the second harmonic (SH) wave $E=\psi\exp(-i\omega_S t + ik_S z)$ generated by a strong, ST localized fundamental pump wave $E=\psi_F\exp(-i\omega_F t + ik_F z)$, $\omega_S=2\omega_F$,  in a nonlinear crystal. In the undepleted pump approximation, the amplification of the SH wave is described by
\begin{equation}\label{SH}
\frac{\partial \psi}{\partial z}= \frac{i}{2k_S}\Delta_\perp \psi - \beta \frac{\partial \psi}{\partial t_F} + i\chi \psi_F^2 e^{-i\Delta k z},
\end{equation}
where $t_F=t-k'_F z$ is the local time for the fundamental wave, $\Delta k=k_S-2k_F$ is the phase mismatch, $\beta=k'_S-k'_F$ is the group delay, and $\chi$ is the nonlinear coefficient. In the above relations the propagation constants are $k_i=k(\omega_i)$, $i=S,F$, and the inverse group velocities are $k'_i=dk(\omega)/d\omega|_i$. The different regimes of SH generation with regard to the fundamental and SH spatiotemporal structure have been studied in detail in \cite{VALIULIS}. In Eq. (\ref{SH}) we assume, as in \cite{CONTI}, a regime in which group delay dominates over other effects such as group velocity dispersion, and that the fundamental wave is a strongly localized wave in space and time that travels substantially undistorted in the medium.

\begin{figure}[!]
\centering
\includegraphics*[height=4cm]{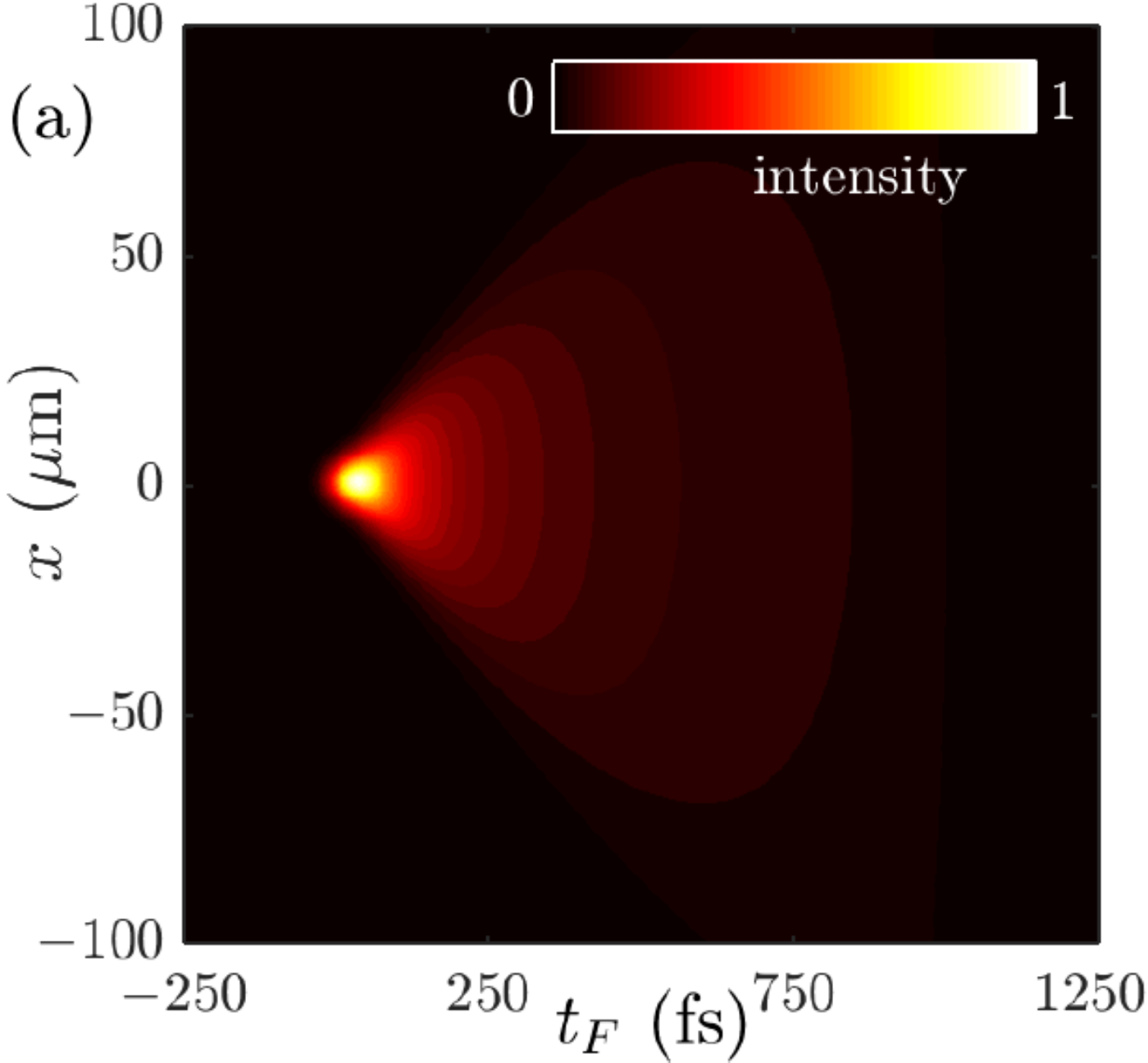}\includegraphics*[height=4cm]{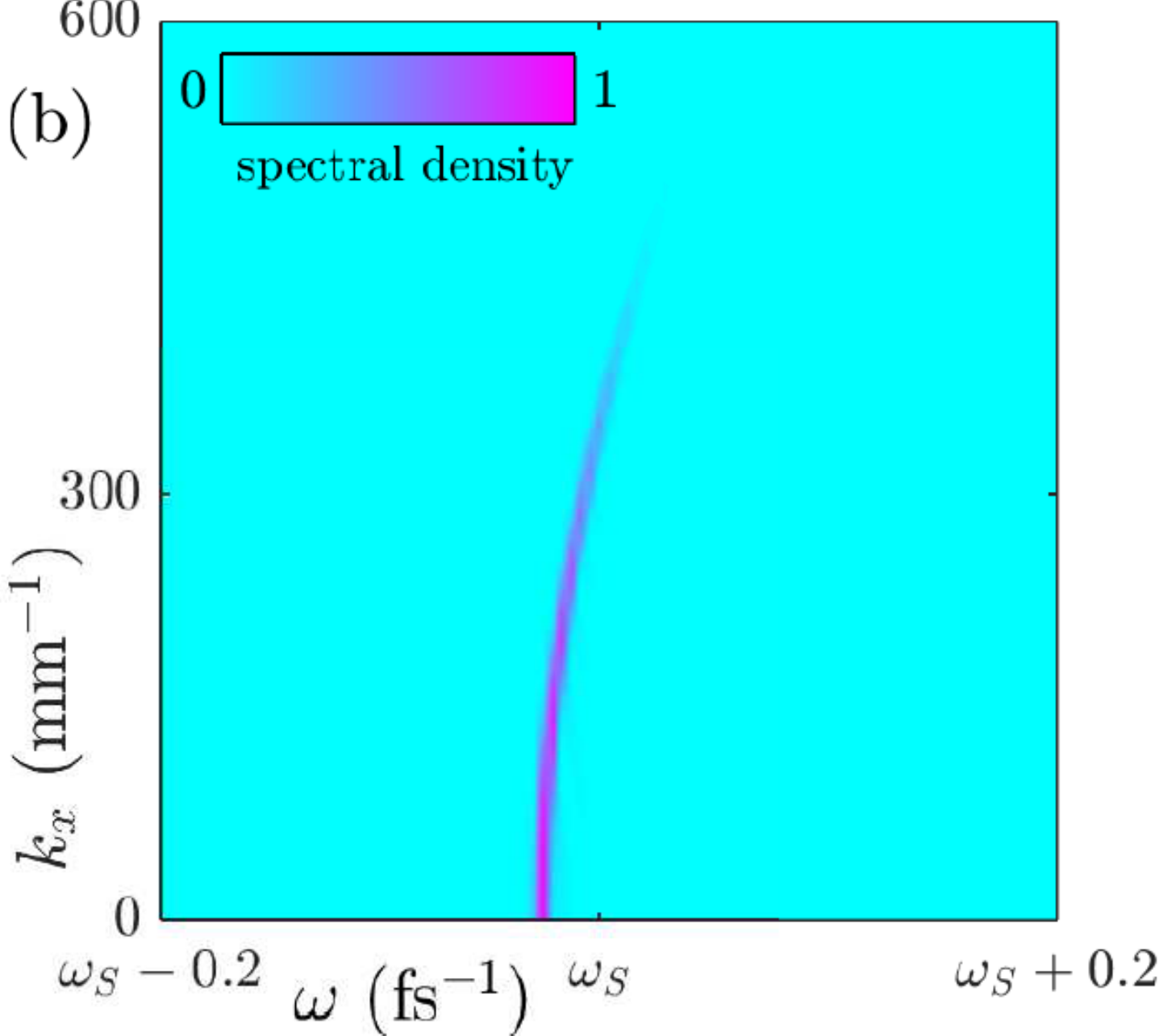}
\includegraphics*[height=3.6cm]{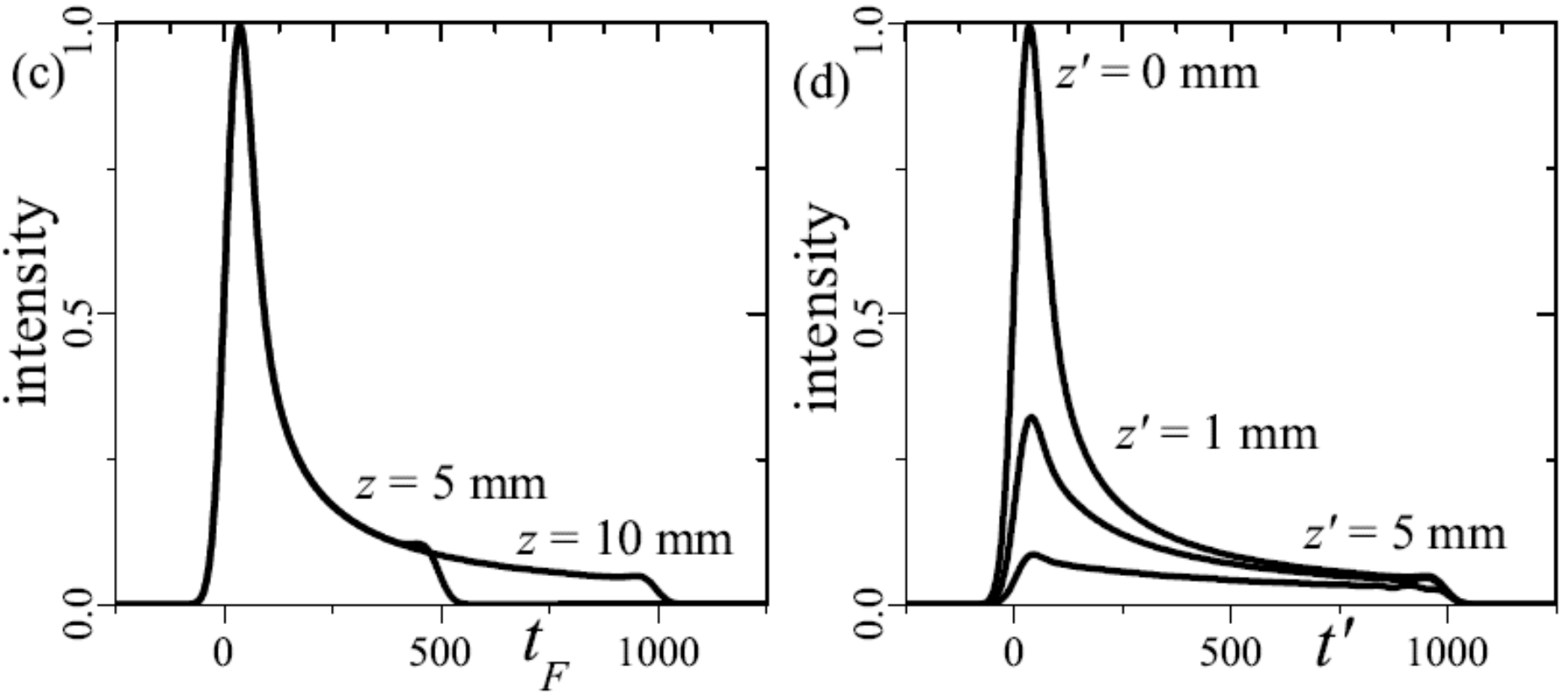}
\includegraphics*[height=4.25cm]{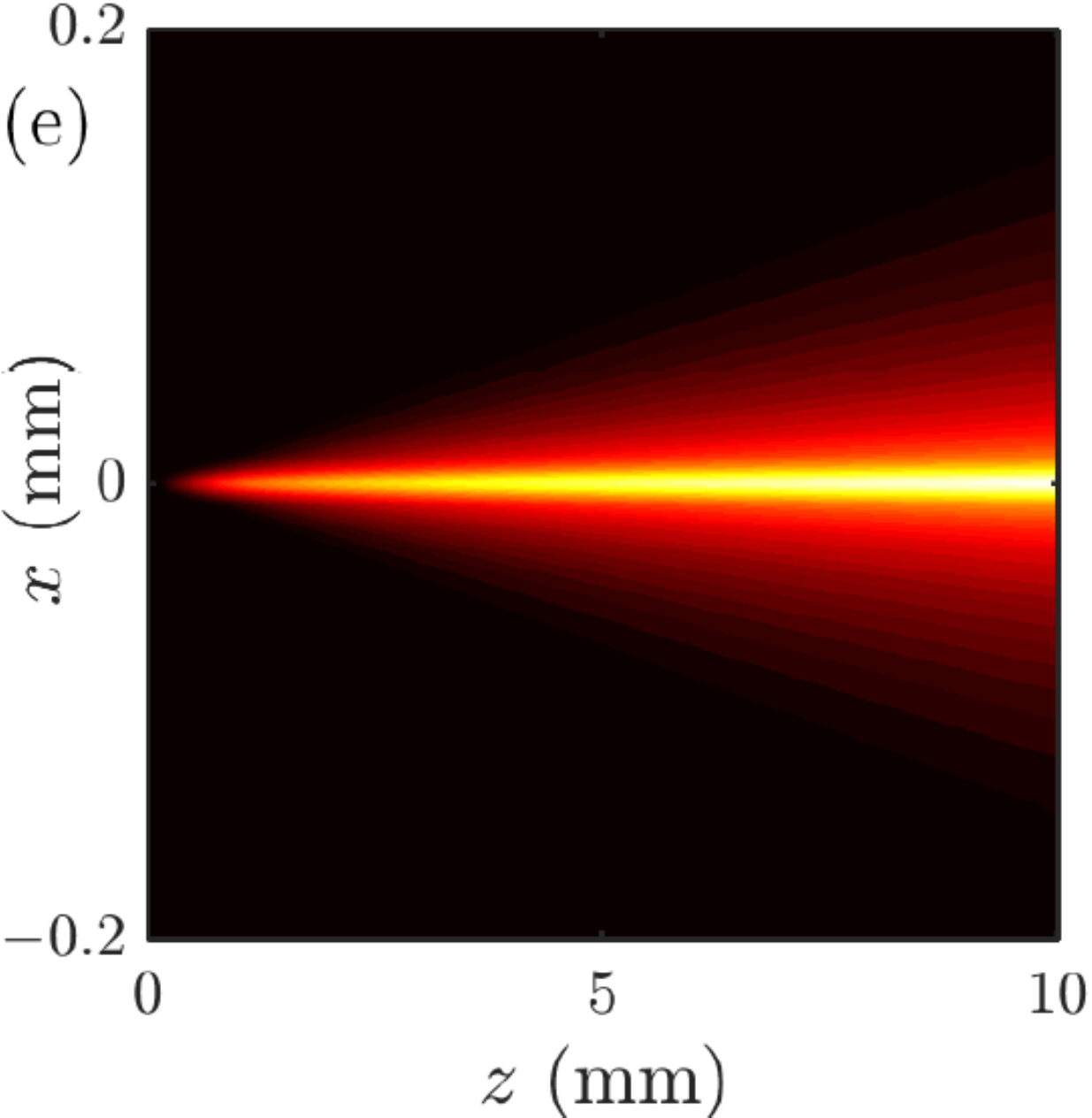}\includegraphics*[height=4.25cm]{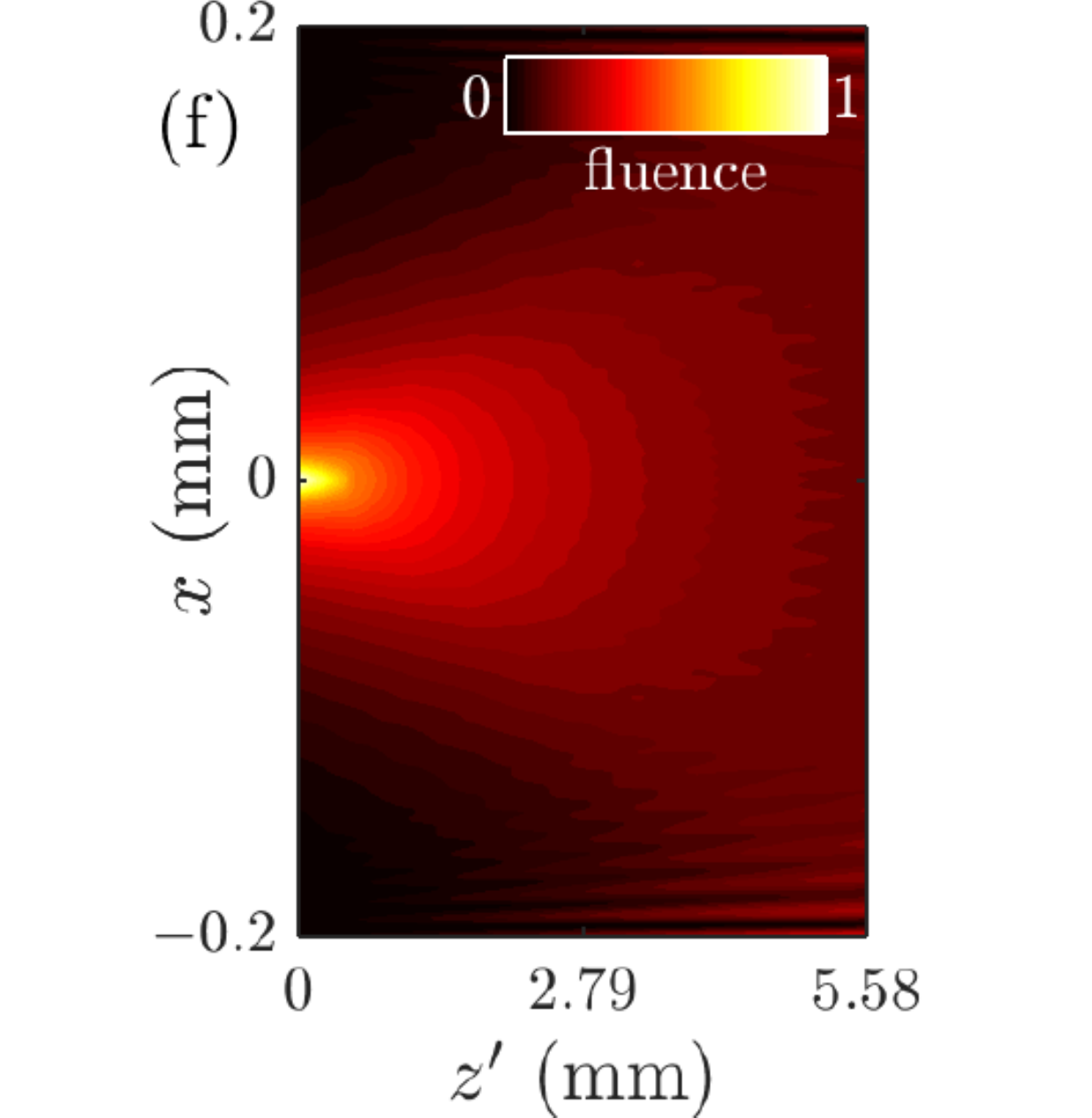}
\caption{\label{Fig5} SH at 530 nm ($\omega_2=3.557$ fs$^{-1}$) generated by a fundamental wave at 1060 nm ($\omega_F=1.778$ fs$^{-1}$) in a nonlinear crystal of length: (a) ST intensity distribution $|\psi|^2$. (b) ST spectral density  $|\hat \psi|^2$, both normalized to their peak values at $z=10$ mm. (c) On-axis pulse shape at $z=5$ mm and $z=10$ mm. The fundamental pulse is the pulsed spatial soliton $\psi_F=\sqrt{I_F}e^{-t_F^2/ T_F^2}\mbox{sech}(x/w_F)e^{ik_{\rm NL}z/2}$, of duration $\Delta t_F=2T_F =141.4$ fs, intensity $I_F=70$ GW/cm$^2$, width $w_F=1/\sqrt{k_Fk_{\rm NL}}=5\,\mu$m, and nonlinear propagation constant $k_{\rm NL}=k_F n_{\rm NL}I_F/n_F$. We took the material parameters $\chi=7\times 10^{-5}$ W$^{-1/2}$, $n_F=1.7774$, $n_S=1.7780$, $n_{NL}=9\times 10^{-15}$ cm$^2$/W, yielding $k_F=n_F(\omega_F/c)=1.0536\times 10^4$, $k_S=n_S(\omega_S/c)=2.1078\times 10^4$ mm$^{-1}$, $\Delta k=k_S-2k_F-k_{\rm NL}= 3$ mm$^{-1}$, and a group mismatch $\beta=100$ fs/mm, compatible with a KTP crystal. (d) On-axis pulse shape at increasing distances $z'$ from the exit face of the crystal at $z=L=10$ mm. (e) Beam fluence in the nonlinear crystal and (f) in the free space beyond the crystal.}
\end{figure}

Figures \ref{Fig5}(a) and (b) show an example of the ST intensity distribution and spectral density, at a distance $z=10$ cm from the input plane of a KTP crystal, of the SH wave at $1060$ nm amplified from noise by a strongly localized fundamental pump wave at $530$ nm, under conditions of small phase mismatch $\Delta k=3$ mm$^{-1}$ for efficient SH generation \cite{CONTI}, and large group mismatch $\beta=100$ fs/mm, as obtained by numerically solving Eq. (\ref{SH}). For simplicity, a two-dimensional or slab geometry is considered ($\Delta_\perp=\partial_{xx})$, in which case the most natural ST localized fundamental wave travelling undistorted is the pulsed spatial soliton $E=\psi_F\exp[-i\omega_F t + i(k_F+k_{NL}/2) z]$, $\psi_F=\sqrt{I_F}e^{-t_F^2/T_F^2}\mbox{sech}(\sqrt{k_Fk_{\rm NL}}x)$, $k_{NL}=k_F n_{\rm NL} I_F/n_F$ of duration $\Delta t_F=2T_F=141.4$ fs and width $w_F=5\,\mu$m (see caption for details). In a three-dimensional geometry, other propagation-invariant pump waves such as multidimensional solitons or nonlinear Bessel beams are also numerically seen produce SH waves of similar characteristics. For different propagation distances $z$, the intensity distribution of the SH wave is the same as in Fig. \ref{Fig5}(a), except that its tail broadens in time, extending at each distance $z$ from the pump location $t_F=0$ to  $t_F=\beta z$, as seen in Fig. \ref{Fig5}(c) for the on-axis intensity. With negative group mismatch, the SH wave would extend from $t_F=\beta z<0$ to $t_F=0$. In Ref. \cite{CONTI}, it was indeed demonstrated the SH wave that tends to be formed is one half of a propagation-invariant, localized wave, which was referred with the generic name of X-wave.

As in Fig. \ref{Fig5}(b), the ST spectrum increasingly concentrates with propagation distance about the parabola $\omega =\omega_S-\Delta k/\beta+|\mathbf{k}_\perp|^2/2k_S\beta$, which corresponds to the precise ST frequency correlations of a TD beam of the carrier frequency $\omega_0=\omega_S-\Delta k/\beta$ (very close to $\omega_S$ in practice). These correlations were shown in \cite{VALIULIS} to arise spontaneously from strongly localized fundamental pump waves as the result of the most efficient amplification of the MPW constituents in the SH wave that are axially phase matched with those of the fundamental wave: The axial propagation constants of the SH frequencies are $k_{S,z}(\omega)=[k_S^2(\omega)-|\mathbf{k}_\perp|^2]^{1/2}\simeq k_S(\omega)-|\mathbf{k}_\perp|^2/2k_S \simeq k_S+k'_S(\omega-\omega_S)-|\mathbf{k}_\perp|^2/2k_S$ in the P\&QM approximations and if group velocity dispersion is not relevant, and the axial propagation constants of the fundamental frequencies are $k_{F,z}(\omega/2)\simeq k_F+k'_F(\omega-\omega_F)/2$. Axial phase matching $\Delta k_z(\omega)\equiv k_{S,z}(\omega)-2k_{F,z}(\omega/2)=0$ yields the parabola $\omega-\omega_S+\Delta k/\beta=|\mathbf{k}_\perp|^2/2k_S\beta$, or introducing the carrier frequency
\begin{equation}\label{OMEGA0}
\omega_0=\omega_S-\frac{\Delta k}{\beta},
\end{equation}
and the detuning $\Omega=\omega-\omega_0$, the parabola $\Omega=|\mathbf{k}_\perp|^2/2k_S\beta$. Free space propagation beyond the output face of the nonlinear crystal does not alter the parabolic spectrum, which is conveniently written as the parabolic ST spectrum of a TD beam in free space as in Eq. (\ref{PARABOLA}), with $k_0=\omega_0/c$, and with a group delay
\begin{equation}\label{ALPHA2}
\alpha =\frac{k_S}{k_0}\beta .
\end{equation}
Since the available temporal frequency bandwidth is $\Delta\Omega =2\Delta \Omega_F\simeq 4/\Delta t_F$, the confocal duration of the TD beam is
\begin{equation}\label{DURATION}
\Delta t\simeq \frac{\Delta t_F}{2},
\end{equation}
and its width $w_0^2=\Delta t/k_0|\alpha|$ is
\begin{equation}\label{WIDTH}
w_0^2\simeq\frac{\Delta t_F}{2k_s|\beta|}.
\end{equation}
All properties of the ideal TD beam are thus fully specified by the pump duration $\Delta t_F$, the group mismatch $\beta$, and the phase mismatch $\Delta k$ in the SH generation process. In the above example, $\omega_0=3.527$ fs$^{-1}$ ($534.5$ nm), $\alpha=179.3$ fs/mm, $\Delta t=70.7$ fs, $w_0=5.79\,\mu$m. We also note that in practice $|\alpha|=(k_S/k_0)|\beta|\ll 1/c$. Then, superluminal (subluminal) TD beams in free space are produced in positively (negatively) group mismatched SH generation, i. e., when the fundamental wave, and therefore the nonlinear polarization source, are superluminal (subluminal) for the SH wave.

Also, if the length of the medium is $L$, the duration of the envelope of the TD beam, once it exists to free space, is fixed to $\Delta t_e =|\beta|L$, which yields, {\it if the TD beam existed in an unbounded free space}, the diffraction-free distance $L_{\rm free}=\Delta t_e/|\alpha|$ i. e.,
\begin{equation}\label{LFREE}
L_{\rm free}=\frac{k_0}{k_S}L,
\end{equation}
and the number of times that the TD beam beats diffraction
\begin{equation}\label{LLFREE}
\frac{L_{\rm free}}{L_R}\simeq \frac{2|\beta|L}{\Delta t_F}.
\end{equation}
Taking a crystal length $L=10$ mm in the above example, we would have $L_{\rm free}=5.58$ mm and $L_{\rm free}/L_R\simeq 14$. Surprisingly, the TD beam does not exhibit any appreciable diffraction-free behavior in the half space beyond the crystal, as illustrated in Fig. \ref{Fig5}(d) by the rapidly decreasing intensity of the on-axis pulse shape at increasing distances $z'$ from the exit facet of the medium. Also, the building-up fluence inside the crystal in Fig. \ref{Fig5}(e), quickly spreads in Fig \ref{Fig5}(f) in free space propagation beyond the crystal. The effective diffraction-free distance in the half space $z'>0$ vanishes because, as it is generated at $z'=0$, the temporal waist of the superluminal (subluminal) ideal TD beam is at the leading (trailing) front of the envelope, traveling in free space at velocity $c$, and therefore the temporal waist and envelope cease immediately to overlap upon propagation in $z'>0$ (they would overlap ``virtually" at $z'<0$).

All the above properties can be confirmed analytically following a similar analysis as in \cite{CONTI}. The solution to Eq. (\ref{SH}) for unseeded SH wave [$\psi(\mathbf{r}_\perp,t_F,0)=0$] is given (apart from a constant amplitude) by
\begin{eqnarray}\label{SHT}
\psi(\mathbf{r}_\perp,t_F,z)&=&f(t_F,z)e^{i\frac{\Delta k}{\beta}t_F} e^{-i\Delta k z}\int d\mathbf{k}_\perp \phi(\mathbf{k}_\perp)\nonumber \\
&\times & e^{-i\frac{|\mathbf{k}_\perp|^2}{2k_S\beta}t_F} e^{i\mathbf{k}_\perp\cdot\mathbf{r}_\perp},
\end{eqnarray}
where $\phi(\mathbf{k}_\perp)=\hat I_F(\mathbf{k}_\perp,\Omega_p)$, $\Omega_p=|\mathbf{k_\perp}|^2/2k_S\beta - \Delta k/\beta$, $\hat I_F(\mathbf{k}_\perp,\Omega)$ is three-dimensional Fourier transform of $\psi_F^2(\mathbf{r}_\perp,t_F)$, and $f(t_F,z)$ is a unit rectangle function between $t_F=0$ and $t_F=\beta z$. In the slab geometry $\mathbf{k}_\perp$ and $\mathbf{r}_\perp$ are to be replaced with $k_x$ and $x$. The integral in Eq. (\ref{SHT}) is in fact a non-diffracting wave traveling at the velocity of the fundamental wave, which is enveloped by $f(t_F,z)$ travelling at the same velocity and widening in time, as described numerically. At the exit plane of a nonlinear medium of length $L$, the envelope is conveniently written in the form
\begin{equation}
\psi(\mathbf{r}_\perp,t,L)=f(t,L)e^{i\phi}\int d\mathbf{k}_\perp\phi(\mathbf{k}_\perp) e^{-i\frac{|\mathbf{k}_\perp|^2}{2k_0\alpha}t} e^{i\mathbf{k}_\perp\cdot\mathbf{r}_\perp} ,
\end{equation}
where $k_0=\omega_0/c$, $\omega_0$ is given by Eq. (\ref{OMEGA0}), $\alpha$ is given by Eq. (\ref{ALPHA2}), $e^{i\phi}=e^{i(k_2-\omega_2-\Delta k)L}$ is a constant phase, and $t=t_F-k'_FL$ is a time with origin at the instant of arrival of the SH wave at the exit plane. Propagation a distance $z'$ in free space according to Eq. (\ref{PWE}) yields
\begin{eqnarray}\label{TDSH}
\psi(\mathbf{r}_\perp,t',z')\!&=&\!f(t',L)\!\int\! d\mathbf{k_\perp}\phi(\mathbf{k}_\perp) e^{-i\frac{|\mathbf{k}_\perp|^2}{2k_0\alpha}(t'+\alpha z')} e^{i\mathbf{k}_\perp\cdot\mathbf{r}_\perp} \nonumber \\
&=&\!f(t',L)\psi_\alpha(\mathbf{r}_\perp,t',z') ,
\end{eqnarray}
where we have omitted the irrelevant constant phase $e^{i\phi}$, and $t'=t-z'/c$ is the usual local time for propagation in free space. Equation (\ref{TDSH}) represents a TD beam with the characteristics described above, and with a luminal envelope $f(t',L)$ spanning from the TD temporal waist $t'=0$ to $t'=\beta L$, which explains the lack of diffraction-free behavior.

\begin{figure}[!]
\centering
\includegraphics*[height=4.1cm]{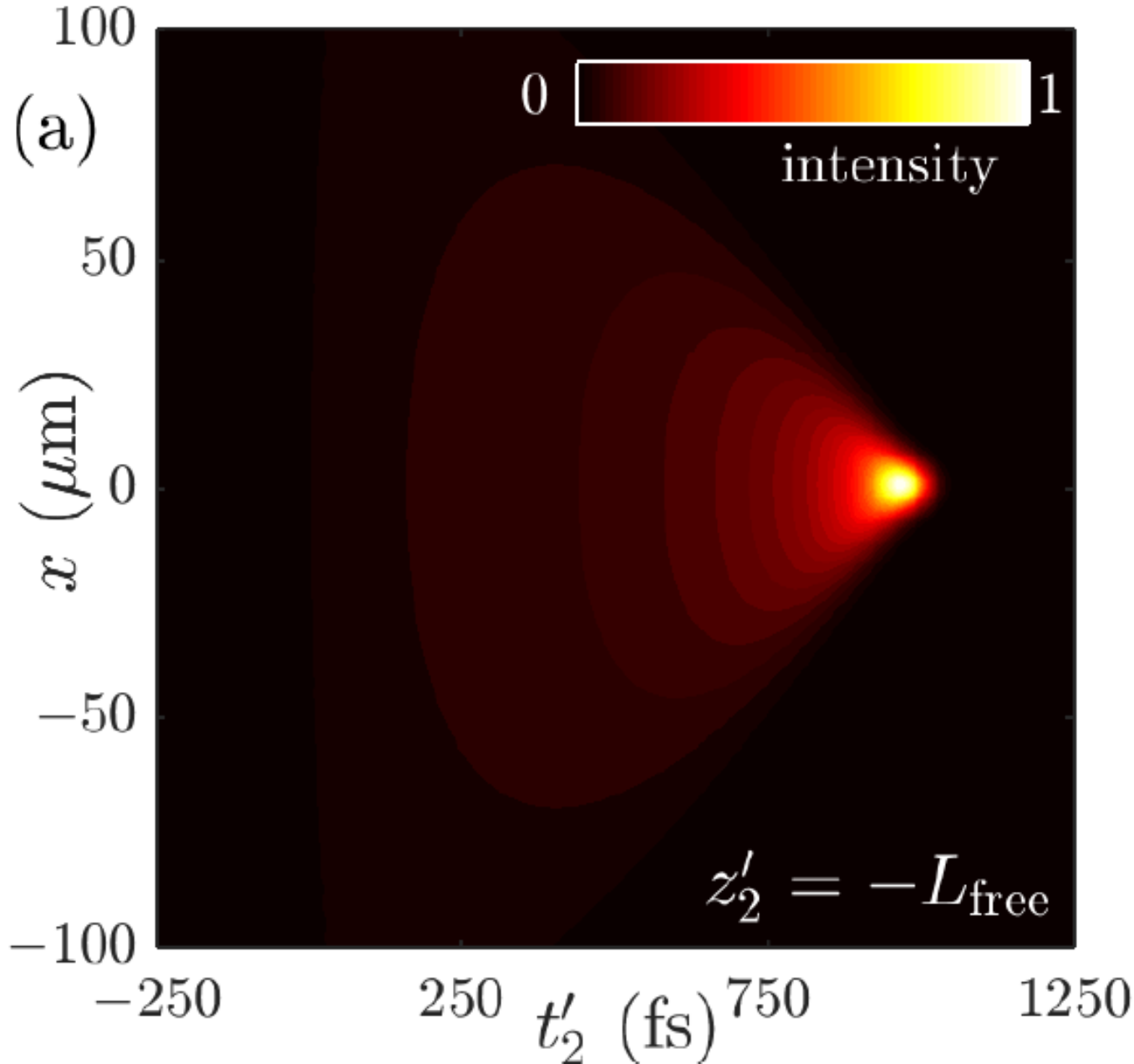}\includegraphics*[height=4.1cm]{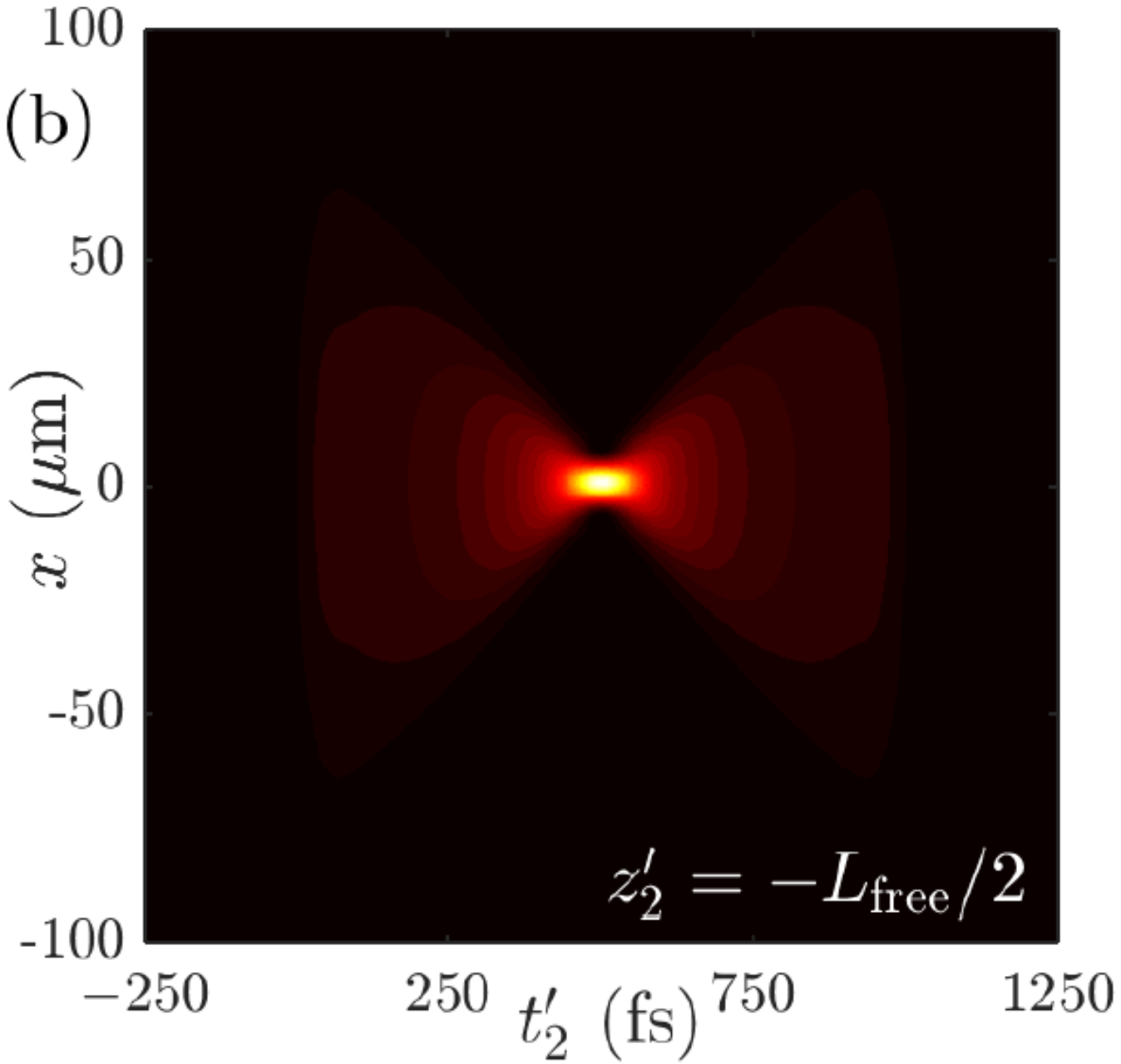}
\includegraphics*[height=3.7cm]{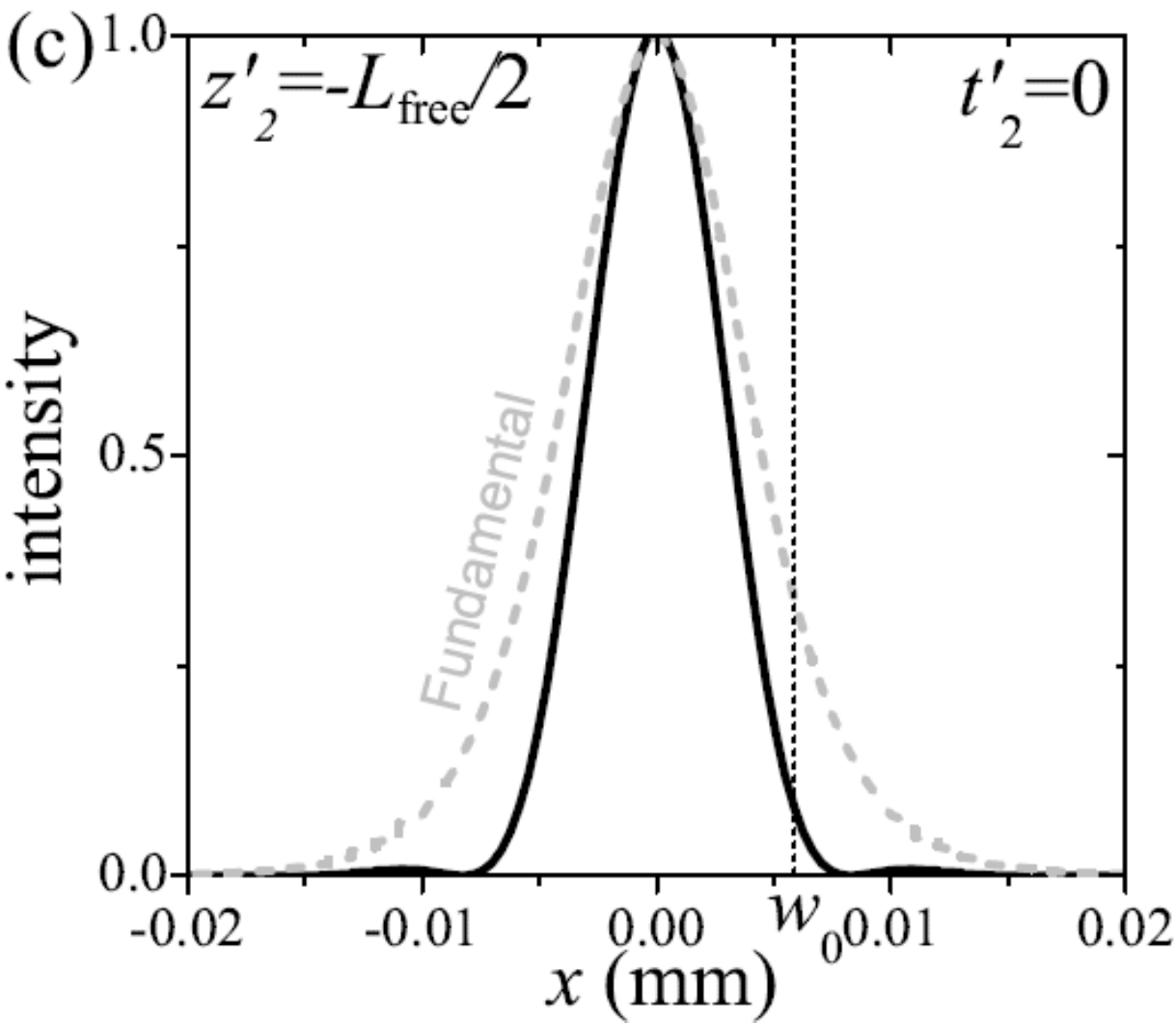}\includegraphics*[height=3.7cm]{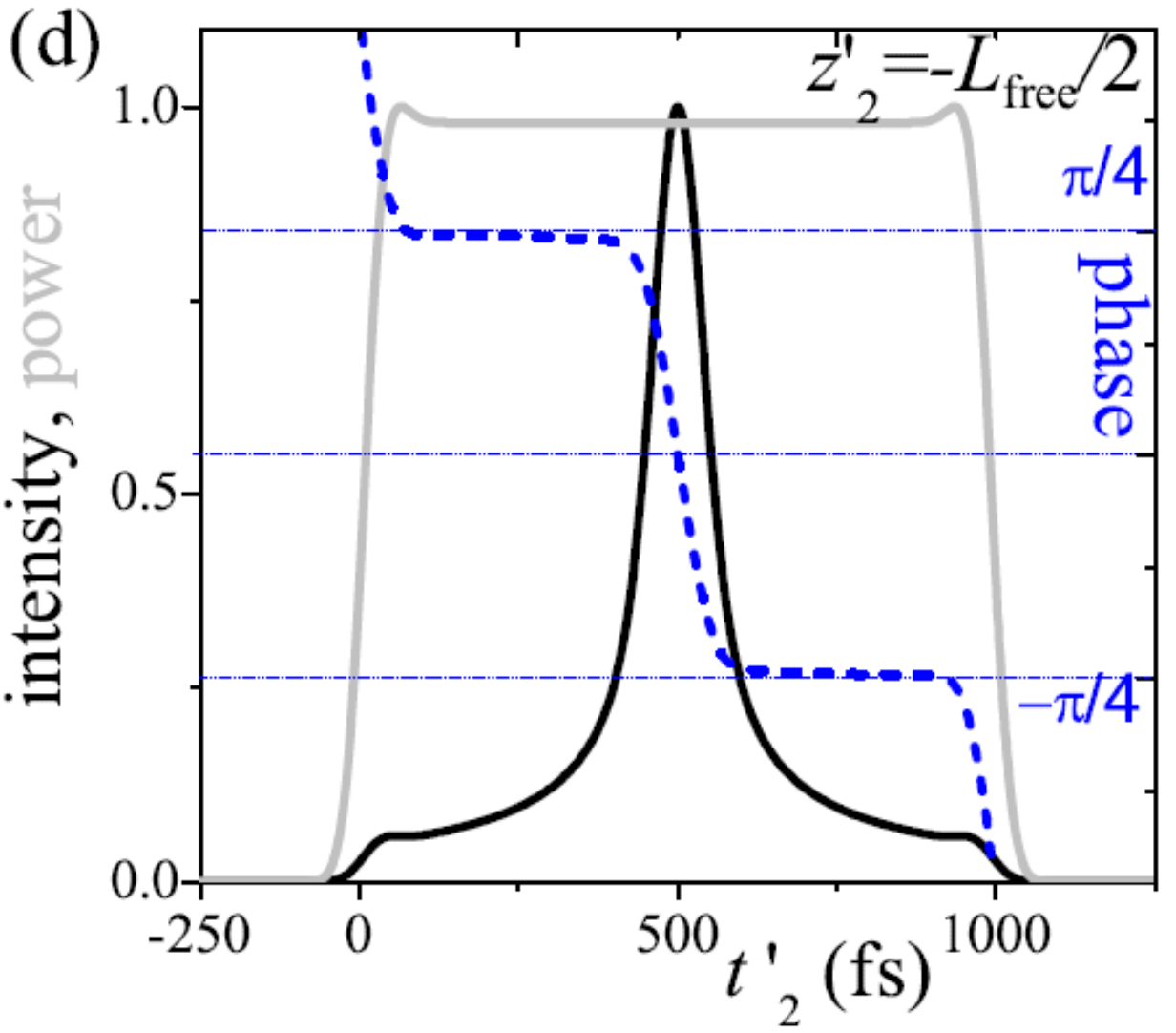}
\includegraphics*[height=3.55cm]{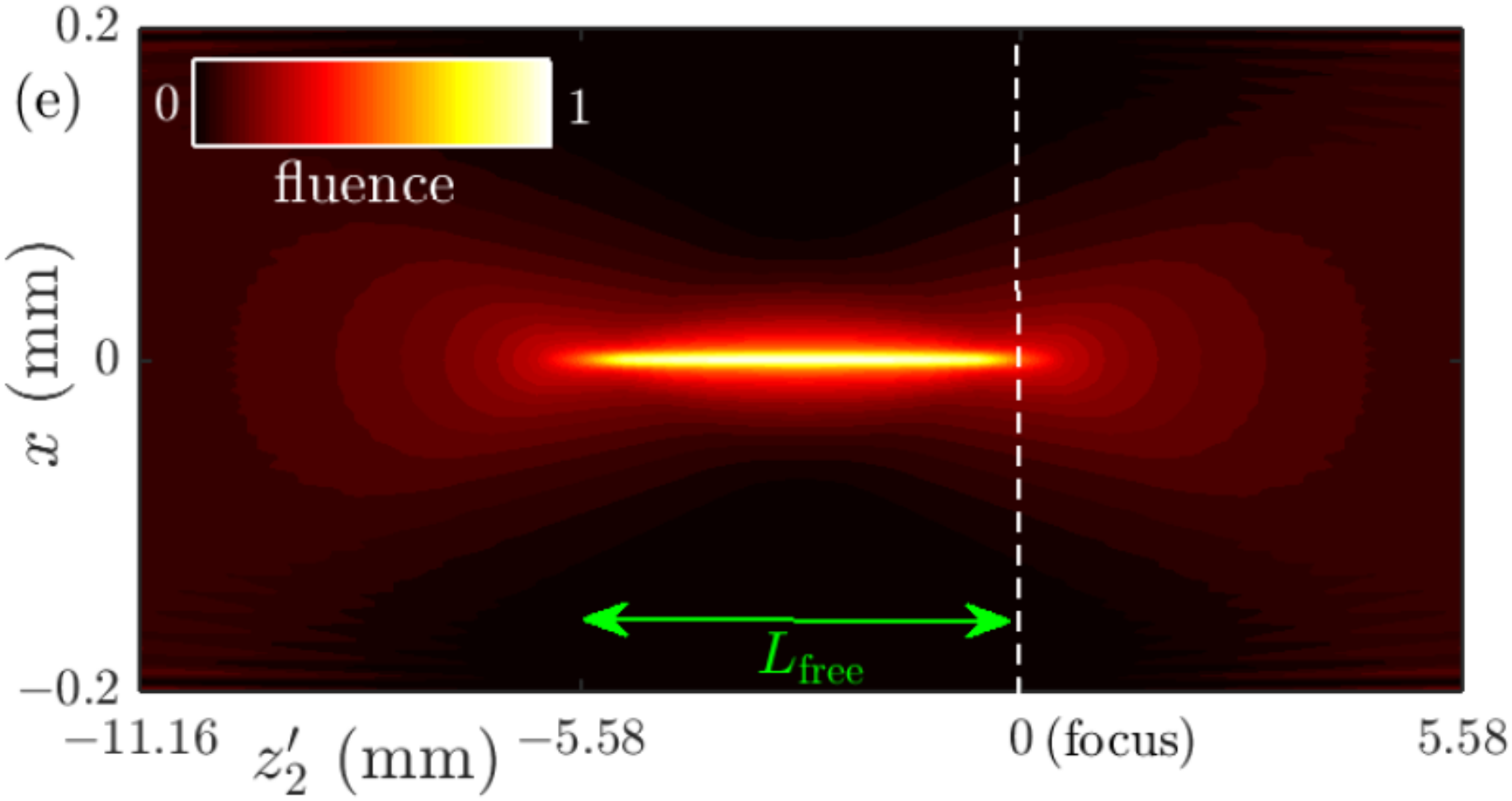}\includegraphics*[height=3.6cm]{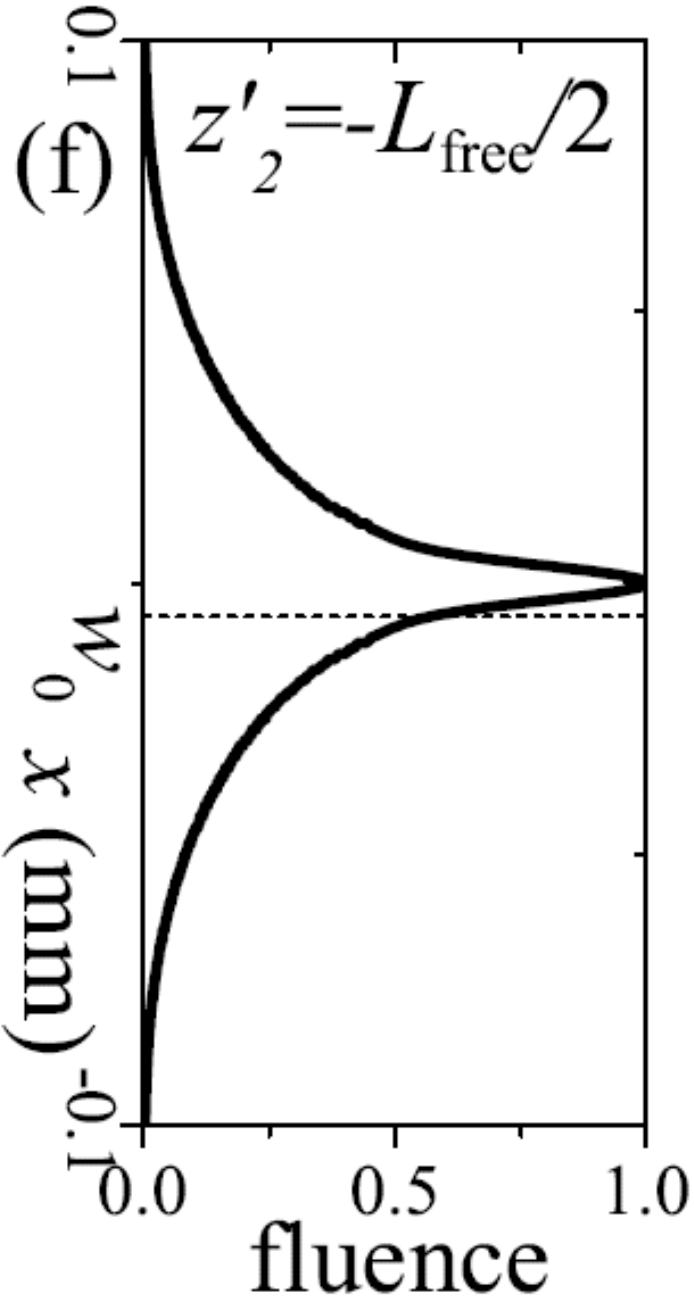}
\caption{\label{Fig6} Image of the TD beam exiting from the nonlinear crystal of Fig. \ref{Fig5} by a $4f$ system of equal focal lengths $f_1=f_2=f$. The origin of the coordinate $z'_2$ is at the back focal plane of the system, and $t'_2=t'-4f/c-z'_2/c$. (a) and (b) ST intensity distribution at the indicated distances, (c) transversal intensity profile at the temporal waist at $z'_2=-L_{\rm free}/2$ , (d) on-axis temporal shape of the intensity (black curve), instantaneous power (gray curve) and phase of the envelope (blue curve) at $z'_2=-L_{\rm free}/2$, showing a temporal Gouy's phase shift about the temporal focus. (d) Fluence distribution in the back focal region and (e) fluence transversal profile at the middle of the needle of light $z'_2=-L_{\rm free}/2$.}
\end{figure}

In order to produce a true needle of light with $L_{\rm free}/L_R=2|\beta|L/\Delta t_F$ determined by the SH generation process, we place the front focal plane of a $4f$ system at the exit plane of the crystal. Figure \ref{Fig6} shows the relevant features of the TD beam formed about the back focal plane of the system with $f_1=f_2=f$ (no magnification), obtained using the numerically simulated SH wave at the exit plane of the crystal of length $L=10$ mm. If $f>L_{\rm free}$, the temporal waist of the TD beam and its envelope start to overlap at a distance $z'_2=-L_{\rm free}$ before the back focus, as seen in the ST intensity profile in Fig. \ref{Fig6}(a), and keep overlapping until the focus at $z'_2=0$, where the ST profile is just the one previously shown in Fig. \ref{Fig5}(a). At $z'_2=-L_{\rm free}/2$, the temporal waist is centered on the envelope, as in Fig. \ref{Fig6}(b). Here we observe all signatures of a beam that indeed diffracts in time: The on-axis intensity within the envelope is seen to decay inversely proportional to time from its temporal waist, as a standard diffracting beam wiht one transversal dimension does axially. The instantaneous power [Fig. \ref{Fig6}(d), gray curve] is constant in time, as it is axially in a diffracting beam. We have also plotted [Fig. \ref{Fig6}(d), dashed curve] the phase to observe the {\it temporal Gouy's phase shift} through the temporal focus by $-\pi/2$ in the one-dimensional geometry. Further the relevant properties of the TD beam are determined by the SH generation process: The width of the transversal intensity profile at the temporal waist $t_2'=0$ [Fig. \ref{Fig6}(c)] is determined by the duration of the fundamental wave and the group delay as given by Eq. (\ref{WIDTH}), the duration $\Delta t$ of the on-axis temporal intensity distribution [Fig. \ref{Fig6}(d), black curve], or temporal depth of focus, is also correctly given by Eq. (\ref{DURATION}), and are therefore related by $\Delta t=k_0|\alpha|w_0^2$ with $\alpha$ given by Eq. (\ref{ALPHA2}). The fluence spatial distribution shown in Fig. \ref{Fig6}(e) and Fig. \ref{Fig6}(f) at its center $z'_2=-L_{\rm free}/2$, forms a needle of light of the length $L_{\rm free}=5.58$ mm, i. e., with $L_{\rm free}/L_R\simeq 14$, as given by Eqs. (\ref{LFREE}) and (\ref{LLFREE}) as functions of the SH generation parameters.

Choosing a magnification different from unity ($f_1\neq f_2$), the TD beam about the back focal plane would be
\begin{eqnarray}
\psi(\mathbf{r}_\perp,t'_2,z'_2)&=&\frac{f_1}{f_2}f(t'_2,L)\!\int\! d\mathbf{k_\perp}\phi\left(-\frac{f_2}{f_1}\mathbf{k}_\perp\right) \nonumber \\ &\times& e^{-i\frac{|\mathbf{k}_\perp|^2}{2k_0\alpha_2}(t'_2+\alpha_2 z'_2)} e^{i\mathbf{k}_\perp\cdot\mathbf{r}_\perp},
\end{eqnarray}
which is a scaled TD as explained in Sec. \ref{TRANSFORMATIONS}, with the same ratio $L_{\rm free}/L_R$ and a group delay $\alpha_2=(f_1^2/f_2^2)\alpha$. In the above example, we would need a magnification $f_1/f_2=13.66$ to produce the abrupt TD beam at infinite speed $\alpha_2=1/c$, or double to produce the back-propagating focus wave mode ($\alpha_2=2/c$).

\section{Conclusion}

As a concluding remark, one of the purposes of this work has been to put the significant and recent advances in the field of diffraction-free waves in closer connection with long-standing knowledge in the same field, both in linear and nonlinear optics. We have settled down the precise meaning and conditions to observe diffraction in time, which are the same as those necessary to realize paraxial and quasi-monochromatic localized waves. Also, in addition to using sophisticated pulse and beam shaping techniques, the spatiotemporal frequency couplings required for the formation of a time-diffracting beam have been shown to arise spontaneously in the nonlinear amplification of waves by strongly localized pump waves. The paraxial and quasi-monochromatic regime of propagation of these localized waves eases enormously the analysis of the behavior of localized waves by optical systems. Putting all these ideas together can help synthesize and efficiently use this fascinating type of waves in their promised applications.

The author acknowledges support from Projects of the Spanish Ministerio de Econom\'{\i}a y Competitividad No. MTM2015-63914-P, and No. FIS2017-87360-P.

\end{document}